\documentclass[12pt,preprint]{article}
\usepackage{graphicx,wrapfig,natbib,epsfig}
\newcommand\plotone[1]{%
 \typeout{Plotone included the file #1}
 \centering
 \leavevmode
 \includegraphics[width={\columnwidth}]{#1}%
}%
\newcommand\plottwo[2]{{%
 \typeout{Plottwo included the files #1 #2}
 \centering
 \leavevmode
 \columnwidth=.45\columnwidth
 \includegraphics[width={\columnwidth}]{#1}%
 \hfil
 \includegraphics[width={\columnwidth}]{#2}%
}}%

\newcommand\aj{{AJ}}%
\newcommand\actaa{{Acta Astron.}}%
\newcommand\araa{{ARA\&A}}%
\newcommand\apj{{ApJ}}%
\newcommand\apjl{{ApJ}}%
\newcommand\apjs{{ApJS}}%
\newcommand\apss{{Ap\&SS}}%
\newcommand\aap{{A\&A}}%
\newcommand\aapr{{A\&A~Rev.}}%
\newcommand\baas{{BAAS}}%
\newcommand\icarus{{Icarus}}%
\newcommand\mnras{{MNRAS}}%
\newcommand\nar{{New A Rev.}}%
\newcommand\pasp{{PASP}}%
\newcommand\ssr{{Space~Sci.~Rev.}}%
\newcommand\nat{{Nature}}%
\newcommand\procspie{{Proc.~SPIE}}%
\let\la=\lesssim            % For Springer A&A compliance...
\let\ga=\gtrsim
\usepackage{amssymb}

\newcommand{\mjupe}{M_{\rm JUP}}
\newcommand{\msini}{\ensuremath{m \sin{i}}}

\newcommand{\mjup}{\ensuremath{\mbox{M}_{\mbox{\tiny Jup}}}}
\newcommand{\mearth}{\ensuremath{\mbox{M}_{\oplus}}}

\def\astrosun {\mbox{$\odot$}}

\newcommand{\Msol}{\ensuremath{M}_{\astrosun}}

\newcommand{\sw}{\ensuremath{\sin \omega_*}}
\newcommand{\cw}{\ensuremath{\cos \omega_*}}
\newcommand{\cO}{\ensuremath{\cos \Omega}}
\newcommand{\sO}{\ensuremath{\sin \Omega}}
\newcommand{\ci}{\ensuremath{\cos i}}
\newcommand{\si}{\ensuremath{\sin i}}
\newcommand{\farcs}{\mbox{\ensuremath{.\!^{\prime\prime}}}}%  % fractional arcsecond symbol: 0.''0
\newcommand\beq{\begin{equation}}
\newcommand\eeq{\end{equation}}

\newcommand\thetae{\theta_{\rm E}}
\def\snr{{\rm S/N}}
\def\e{{\rm E}}
\def\thetae{\theta_\e}
\def\drel{d_{\rm rel}}
\def\la{\lesssim}
\def\ga{\gtrsim}
\let\oldbibliography\thebibliography
\renewcommand{\thebibliography}[1]{%
  \oldbibliography{#1}%
  \setlength{\parskip}{0.0pt}%
  \setlength{\itemsep}{0.0pt}%
  }
\title{Exoplanet Detection Methods}
\author{Jason T. Wright, B.\ Scott Gaudi}

\begin{document}
\maketitle

\begin{abstract}
This chapter reviews various methods of detecting planetary
companions to stars from an observational perspective, focusing on
radial velocities, astrometry, direct imaging, transits, and
gravitational microlensing.  For each method, this chapter first derives or
summarizes the basic observable phenomena that are used to infer the
existence of planetary companions, as well as the physical properties of the
planets and host stars that can be derived from the measurement of
these signals.  This chapter then outlines the general experimental requirements
to robustly detect the signals using each method, by comparing their
magnitude to the typical sources of measurement uncertainty.  This chapter goes on
to compare the various methods to each other by outlining the regions
of planet and host star parameter space where each method is most
sensitive, stressing the complementarity of the ensemble of the
methods at our disposal.  Finally, there is a brief review of the history of
the young exoplanet field, from the first detections to current
state-of-the-art surveys for rocky worlds.
\end{abstract}  
\tableofcontents
\pagestyle{empty}
\markboth{{\sc Wright \& Gaudi}}{{\sc Exoplanet Detection Methods}}
%\newpage
\pagestyle{plain}
%\setcounter{page}{1}

%\clearpage
%\maketitle

\section{Basic Principles of Planet Detection}

This chapter begins by reviewing the basic phenomena that 
are used to detect planetary companions to stars using various
methods, namely radial velocities, astrometry, transits, timing,
and gravitational microlensing.  It derives the generic observables for
each method from the physical parameters of the planet/star system. 
These then determine the physical parameters that can be inferred
from the planet/star system for the general case.

Notation with subscripts $*$ and $p$ refer to
quantities for the star and planet, respectively.  Therefore, a star has mass $M_*$, radius $R_*$, mean density $\rho_*$,
surface gravity $g_*$, and effective temperature $T_*$,
and is orbited by a planet of mass $M_p$, radius $R_p$, density $\rho_p$,
temperature $T_p$, and surface gravity $g_p$.   The orbit
has a semimajor axis $a$, period $P$, and eccentricity $e$. 

\subsection{Spectroscopic Binaries and Orbital Elements}

Exoplanet detection is essentially the extreme limit of binary star
characterization, and so it is unsurprising that the terminology and
formalism of planetary orbits derives from that of binaries.  

Conservation of momentum requires that as a planet
orbits a distant star, the star executes a smaller, opposite orbit about their
common center of mass.  The size (and velocity) of this orbit is
smaller than that of the planet by a factor of the ratio of their
masses. The component of this motion along the line of sight to the
Earth can, in principle, be detected as a variable radial velocity.
The mass of the exoplanet can be calculated from the magnitude of the
radial velocity (RV) or astrometric variations and from the mass of
the star, determined from stellar models and spectroscopy or astrometry.  

\begin{figure}
\plotone{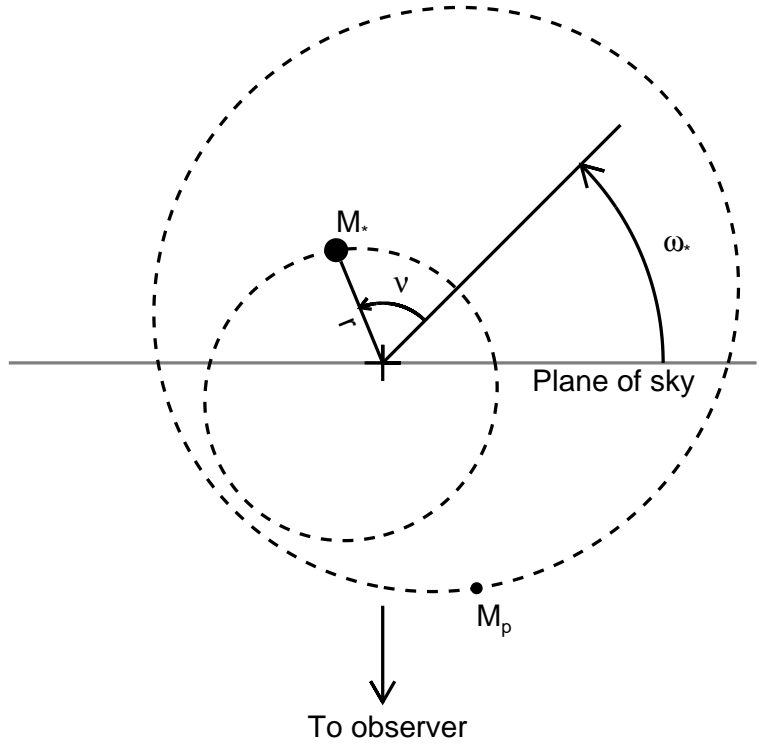}
\caption{Elements describing orbital motion in a binary with respect
  to the center of mass (cross).  The argument of periastron
$\omega$ is measured from the ascending (receding) node, and the
true anomaly $\nu$ is measured with respect to the periastron.  Both angles increase along the
direction of the star's motion in the plane of the orbit.  The
longitude of the periastron of the star $\omega_*$ is indicated.  At a given
time in the orbit, the true anomalies of the planet and star are equal,
where as the longitude
of periastron of the planet is related to that of the star by $\omega_p=\omega_* + \pi$.  In
Doppler planet detection, the orbital elements of the star are conventionally
reported, from which the orbital elements of the planet can be inferred. \label{orbvar}}
\end{figure}

Two mutually orbiting bodies revolve in ellipses about a common center of mass, the origin of our
coordinate system.  Orbital angles in the plane of the bodies'
mutual orbit are measured with respect to the line of nodes, formed by the intersection of the orbital plane with the
plane of the sky (i.e. the plane perpendicular to a line connecting the
observer to the system's center of mass).  The position of
this line on the sky has angle $\Omega$, representing the
position angle (measured east of north)  of the ascending (receding)
node, where the star (and planet) cross the plane of the sky moving
away from Earth. Figure~\ref{orbvar} illustrates the other orbital elements in
the problem.  As indicated, the orientation of the each orbital ellipse with respect to the plane
of the sky is specified by the longitude of periastron, $\omega$, which is the angle between
the periastron \footnote{Periastron marks the point where the two bodies
  have their closest approach.} and the ascending node along the orbit in the direction of the motion of
the body.  Since the orbit of the 
star is a reflection about the origin of the orbit of the exoplanet,
the orbital parameters of the planet are identical to 
that of the star except that the longitudes of periastron differ by $\pi$: $\omega_p=\omega_*+\pi$.

The physical size of the ellipse, given by the semi-major 
axis, $a$, is set by Newton's modification of Kepler's third law of
planetary motion: 
\begin{equation}
  P^2 = \frac{4 \pi^2 }{G (M_*+M_p)} a^3
\end{equation}

The semimajor axis is $a=a_*+a_p$, where $a_*$ and $a_p$ are the semi-major axes of 
the two bodies' orbits with respect to the center of mass, given by,

\begin{equation}
  a_* = \frac{M_p}{M_*+M_p} a;\qquad  a_p= \frac{M_*}{M_*+M_p} a
\end{equation}

\begin{figure}
\plotone{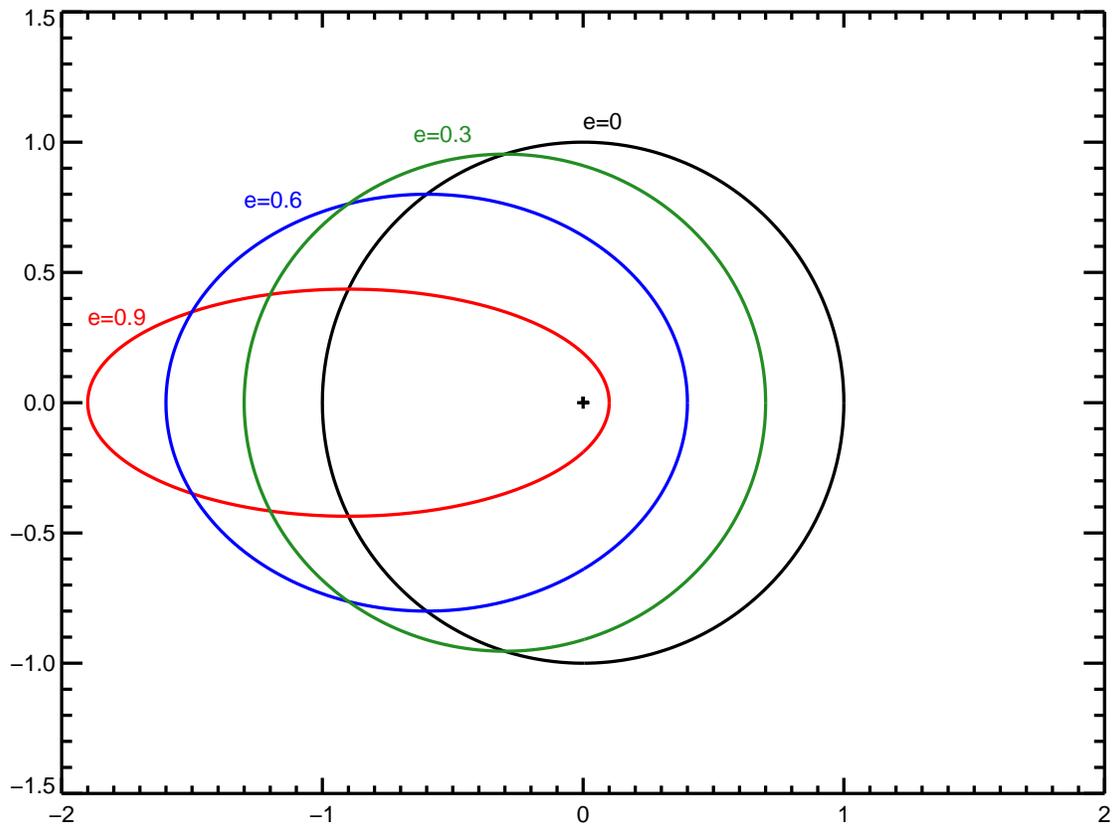}
\caption{Shape of various eccentric orbits in the orbital plane.  A
  handful of exoplanets with eccentricities above 0.9 
 have been detected.\label{eccs}}
\end{figure}

The position of either body in its orbit about the origin can be
expressed in polar coordinates $(r,\nu)$, where $\nu$ is the true anomaly, 
the angle between the location of the object and the
periastron.  The separation between the star and planet is given by
\begin{equation}
 r(1+e\cos\nu) = a (1-e^2) 
\end{equation}
where $e$ is called the eccentricity of the orbit, and has the
domain $[0,1)$ for bound orbits.  The observed eccentricities of exoplanets are quite varied:
eccentricities above 0.9 have been seen in a few cases, and
eccentricities above 0.3 are common, at least for Jovian exoplanets \citet{EOD}.
Figure~\ref{eccs} illustrates the physical shape of such orbits.

Practical computation of a body's position in its orbit with time is
usually performed through the intermediate variable $E$, called the eccentric
anomaly.  $E$ is related to the time since periastron passage $T_0$ through
the mean anomaly, $M$: 

\begin{equation}
M = \frac{2 \pi (t-T_0)}{P} = E - e \sin E
\end{equation}.

\noindent and allows the computation of $\nu$ through the relation

\begin{equation}
 \tan{\frac{\nu}{2}} = \sqrt{\frac{1+e}{1-e}} \tan{\frac{E}{2}}
\end{equation}

\noindent The eccentric anomaly is also useful because it is simply related to $r$:

\begin{equation}
E = \arccos{\frac{1-r/a}{e}}
\end{equation}

\subsection{Radial Velocities}\label{sec:rv}
The radial reflex motion of a star in response to an orbiting planet
can be measured through precise Doppler measurement, and this motion
reveals the period, distance and shape of the orbit, and provides
information about the orbiting planet's mass.  (The treatment of RV
and astrometric measurement below follows \citet{Wright09b}).

Six parameters determine the functional form of the periodic radial
velocity variations and thus the observables in a spectroscopic orbit
of the star: $P$, $K$, $e$, $\omega_*$,
$T_0$, and $\gamma$ (it is convention in the Doppler-detection literature to refer to
$\omega$ without its $*$ subscript, but it is standard to report the
star's argument of periastron, not the planet's).

\begin{equation}
V_r = K [ \cos(\nu+\omega_*) + e \cos\omega_* ] + \gamma
\end{equation}

\noindent with $\nu$ related to $P$, $e$, and $T_0$ through $E$.  The semi-amplitude of
the signal in units of velocity is $K$ (the peak-to-trough RV variation
is $2K$).  The bulk velocity of the center of mass of the system is
given by $\gamma$.

For circular orbits $e=0$, there is no periastron approach, and so $T_0$ and
$\omega_*$ are formally undefined;  in such cases a nominal value of
$\omega_*$ (such as 0 or $\pi/2$) sets $T_0$ (alternatively, one can
specify the value of one of the angles at a given epoch).

\begin{figure}
\plotone{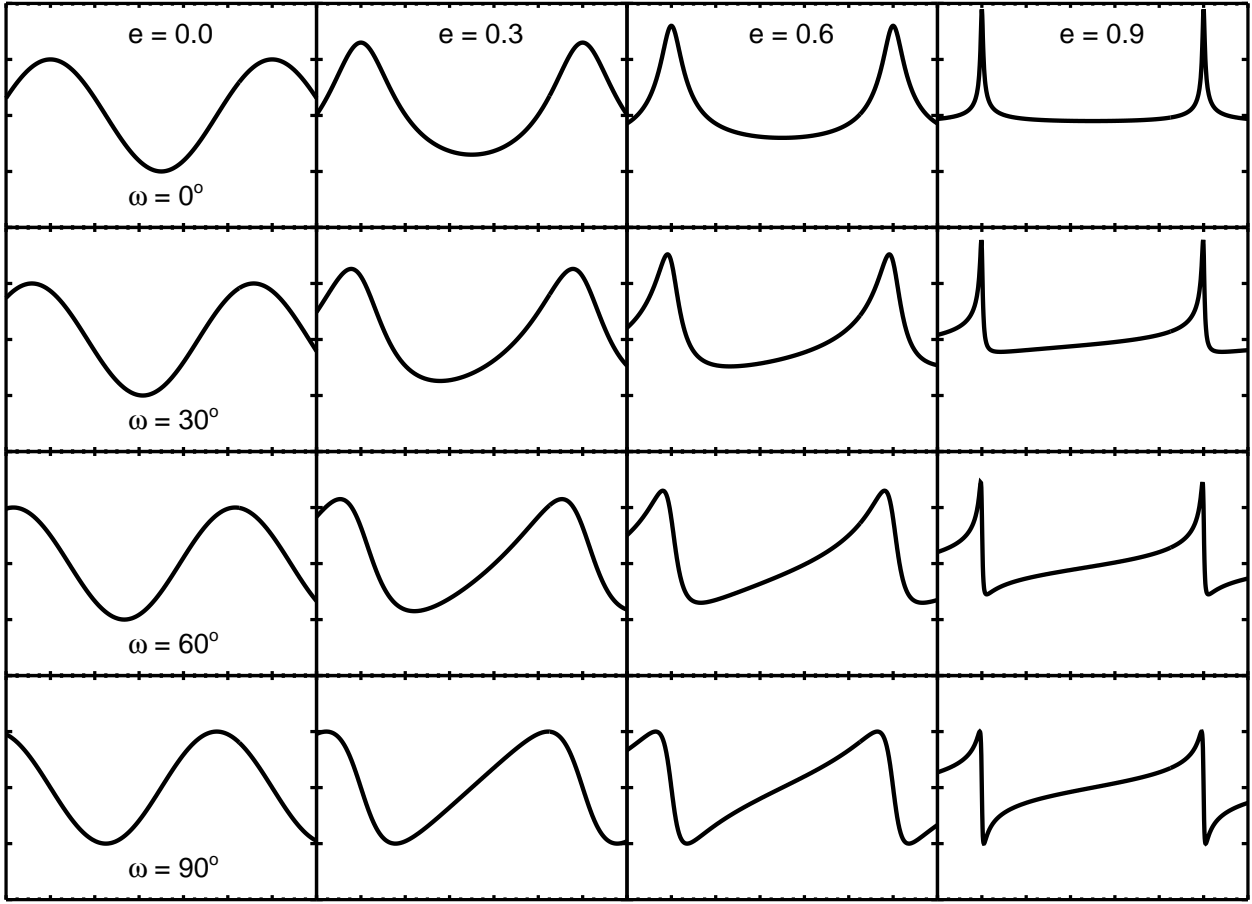}
\caption{The effects of $e$ and $\omega_*$ on on radial velocity
  curves.  These curves have been scaled to unit $K$ and common $P$
  and $T_0$.
  Each column shows curves of constant $e$ and each row curves of
  constant $\omega_*$ as indicated.  Other quadrants of $\omega_*$ yield
  reflections of these curves.\label{rvs}}
\end{figure}

In short, the variables $P$, $T_0$, and $K$ respectively set the period, phase, and
amplitude of an RV curve, while the variables $\omega_*$ and $e$
determine the shape of the radial velocity signature of an orbiting
companion, as shown in Figure~\ref{rvs}.   Characterization of the
orbits of single unseen companions, such as exoplanets, is ultimately
an exercise in fitting observed radial velocities to the family of curves in 
Figure~\ref{rvs} to determine the six orbital parameters.

Two additional orbital parameters complete the description of a planet's
orbit:  the inclination of the orbit, $i$, which determines the angle
between the orbital plane and the plane of the sky, such that $i=0$
corresponds to a face-on, counter-clockwise orbit, and $\Omega$, the longitude
of the ascending node.  These parameters
cannot be determined with radial velocity observations alone, and 
can only be measured through astrometry, where the
angular displacement of the star on the sky is directly measured.

The effect of the inclination of the orbit is to reduce the radial
component of the velocity of the star by $\sin i$.  The fundamental
observable of a spectroscopic binary which constrains
the physical properties of system is thus
\begin{equation}
\label{defmf}
\frac {P K^3  (1-e^2)^{\frac{3}{2}}}{2\pi G}= \frac{M_p^3\sin^3i}{(M_p+M_*)^2}
\end{equation}
where $G$ is Newton's gravitational constant.  The right hand side of
this equation is known as the {\it mass function} of the 
system.  For exoplanets where $M_*$ can be estimated from stellar models,
the minimum value for $M_p$ (i.e. its value for $\sin i = 1$, or an edge-on
orbit) is called the ``minimum mass'' of the planet, and is usually
labeled ``$M_p\sin i$'' for succinctness (since when $M_p \ll M_*$ its small
correction to the denominator is negligible, though not ignored).  The
true mass of the detected exoplanet is thus higher by a factor of
$1/\sin i$, which has a typical (median) value of 1.15 for randomly
oriented orbits (all other things being equal).

\subsection{Astrometry}

Plane-of-sky variations in a star's position provide both redundant
and complementary information to radial velocities, yielding the true
inclination and orientation of a planetary orbit.  Astrometry of the orbits of
well-separated binary stars of similar magnitude is a matter of
careful instrument calibration to precisely measure the separation and
position angle between the stars. For exoplanet detection, the problem
is to detect the motions around a star about an unseen companion with
respect to a set of (presumably) stable background stars. 

For an orbit with semimajor axis $a_*$ of a star at distance $d$ from
the Earth, producing an astrometric signal of semi-amplitude
$\theta_*=a_*/d$, astrometric orbits can be described in terms of the
Thiele-Innes constants 
\begin{eqnarray}
A &=& \theta_*(\phantom{-}\cO \cw - \sO\sw\ci) \\ 
B &=& \theta_*(\phantom{-}\sO \cw + \cO\sw\ci)\\
F &=& \theta_*(-\cO \sw - \sO\cw\ci)\\
G &=& \theta_*(-\sO \sw + \cO\cw\ci)\\
C &=& \theta_*\sw \si \\
H &=& \theta_*\cw \si
\end{eqnarray}
which can be quickly computed using rotation matrices:
\begin{equation}
\label{defTI}
\left[ 
  \begin{array}{ccc}
A & B & C\\
F & G & H\\
\theta_*\sin i\sin \Omega & -\theta_*\sin i \cos \Omega & \theta_*\cos i
  \end{array}
 \right] = \theta_* R_z(\omega_*) R_x(i) R_z(\Omega)
\end{equation}
where $R$ is the 3-D rotation matrix, given in the case of the
$z$-axis by
\begin{eqnarray}
\label{defr}
R_z(\Omega) = \left[ 
 \begin{array}{ccc}
   \phantom{-}\cos \Omega & \phantom{-} \sin \Omega & \phantom{in}0\phantom{-} \\ 
   -\sin \Omega & \phantom{-}\cos \Omega & \phantom{in}0\phantom{-} \\
   0 & 0 & \phantom{in}1\phantom{-}
   \end{array}
\right]
\end{eqnarray}

The Thiele-Innes constants are related back to Keplerian orbital elements with the relations:
\begin{eqnarray}
\label{sumom}
\tan (\omega_*+\Omega) &=& \phantom{-(}\frac{B-F}{A+G}\\
\label{difom}
\tan (\omega_*-\Omega) &=& \frac{-(B+F)}{\phantom{-(}A-G\phantom{)}}\\
\tan^2\left(\frac{i}{2}\right) &=&
\frac{(A-G)\cos(\omega_*+\Omega)}{(A+G)\cos(\omega_*-\Omega)}\
\end{eqnarray}
\begin{equation}
\begin{array}{rl}
\theta_* =&(A\cos\omega_*-F\sin\omega_*)\cos\Omega-\\
&(A\sin\omega_*+F\cos\omega_*)\sin\Omega\sec i
\end{array}
\end{equation}
\begin{equation}
\theta_*^2 = A^2+B^2+C^2=F^2+G^2+H^2
\end{equation}
The quantities $\omega_*$ and $\Omega$  have a $\pm \pi$
ambiguity that is resolved with radial velocity measurements, without
which convention dictates that we choose $\Omega < \pi$. 

The $C$ and $H$ constants are related the radial component of the
motion.  These constants can be combined with the elliptical
rectangular coordinates, defined as 
\begin{eqnarray}
\label{defx}
X & = & \cos E - e\\
\label{defy}
Y & = & \sqrt{1-e^2} \sin E
\end{eqnarray}
\noindent to describe the astrometric
displacements of a star in the North ($\Delta \delta$) and East
($\Delta \alpha \cos \delta$) directions:  
\begin{equation}
\begin{array}{rcl}
\Delta \delta &=& A X + F Y\\
\Delta \alpha \cos \delta&=& B X + G Y
\end{array}
\end{equation}
\noindent and the magnitude of the astrometric offset from the
apparent center of mass is (for small offsets) $\Delta \theta_* \equiv [\Delta \delta^2 +
(\Delta \alpha \cos \delta)^2]^{1/2}$.  In practice, astrometric motions are small perturbations on the much
larger parallactic and proper motions.

\subsection{Imaging}

The direct detection of planets is the most conceptually
straightforward method of detection: essentially one seeks simply
to directly detect photons from the exoplanet, resolved from those of
the parent star.  Although the emission of exoplanets is indeed quite
faint, it is generally the problem of detecting this emission in the proximity of
the much brighter stellar source that presents the most severe practical
obstacle to direct detection.  The disentangling of stellar and planetary photons is an imperfect
process that is easiest at wider separations.  The efficiency of this
disentangling ultimately determines the detection thresholds of an
instrument. Therefore, the most important parameters
of the exoplanet for determining the difficulty of direct detection are
the planet/star flux ratio $f_p$ and the angular separation between
the planet and star.  Typically, contrast limits worsen at smaller angular
separations.  

The angular separation of the planet and star on the sky is given by
\begin{equation}
\Delta \theta = r_\perp / d
\end{equation}
\noindent where $r_\perp$ is
the projected separation of the planet from the star, and $d$ is the distance
to the system.  By definition, if $d$ is in parsecs and $r_\perp$ in AU, then
$\theta$ is in arcseconds.  In general, $\Delta \theta = (1+M_*/M_p)\Delta \theta_* = 
(1+M_*/M_p)\sqrt{(BX+GY)^2+(AX+FY)^2}$. 
For circular orbits, this reduces to $r_\perp = a (\cos^2\beta + \sin^2 \beta \cos^2 i)^{1/2}$,
where $\beta=\nu+\omega_p$ is the angle between the position of the planet along its orbit relative to the 
ascending node.
Planets typically orbit stars at distances from hundredths to hundreds of AU.
For a hypothetical giant planet
orbiting 5 AU from a nearby star sitting at 50 pc, this corresponds to
a maximal angular separation of 100 mas.  

The emission from an exoplanet can generally be separated into two
sources: stellar emission reflected by the planet surface and/or
atmosphere, and thermal emission from the planet.  Thermal emission
can be due to either ``intrinsic'' thermal emission (e.g. the fossil heat of formation), 
or thermal emission from reprocessed stellar luminosity. 
Exoplanets may also produce some non-thermal emission, which we will not consider
here.

The reflected light
will have a spectrum that is broadly similar to that of the star, with
additional features arising from the planetary surface and/or
atmosphere.  Therefore, for solar-type stars, this reflected emission generally
peaks at optical wavelengths.  The monochromatic planet/star flux
ratio for reflected light can generally be written (e.g.,
\citealt{seager2010}),
\beq
f_{{\rm ref},\lambda} = A_{g,\lambda} \left(\frac{R_p}{a}\right)^2 \Phi_{{\rm ref},\lambda}(\alpha)
\label{eqn:frref}
\eeq
where $A_{g,\lambda}$ is the monochromatic geometric albedo, 
and $\Phi_{{\rm ref},\lambda}$ is the reflected light phase curve, which depends
on the planetary phase angle $\alpha$ (the star-planet-observer
angle) and the wavelength $\lambda$.  The geometric albedo is defined as the ratio
of the flux emitted from the planet at $\alpha=0$ relative to that of 
a perfectly and isotropically scattering uniform disk of equal solid angle. 
For a circular orbit, $\cos\alpha=\sin\beta \sin i$.

Assuming that the thermal emission from the planet has a roughly blackbody spectrum,
the flux ratio is
\beq
f_{{\rm therm},\lambda} = \left(\frac{R_p}{R_*}\right)^2 \frac{B_\lambda(T_p)}{B_\lambda(T_*)} \Phi_{{\rm therm},\lambda}(\alpha)
\rightarrow \left(\frac{R_p}{R_*}\right)^2 \frac{T_p}{T_*} \Phi_{{\rm therm},\lambda}(\alpha),
\label{eqn:frthermal}
\eeq
where $\Phi_{{\rm therm},\lambda}$ is the monochromatic thermal phase curve.  
For observations in the Raleigh-Jeans tail of the blackbody, $\lambda \gg hc/(k_b T)$, and
thus $B_\lambda(T) \propto T$, yielding the limit shown in Equation \ref{eqn:frthermal}. 
If the planet is in thermal equilibrium with the stellar radiation,
then $T_p=T_{{\rm eq}}$ and 
\beq
\frac{T_{{\rm eq}}}{T_*}= \left(\frac{R_*}{a}\right)^{1/2} [f(1-A_B)]^{1/4},
\label{eqn:teq}
\eeq
where $A_B$ is the Bond albedo, the fraction of the total energy incident on the planet
that is not absorbed, and $f$ accounts the fraction of the entire planet surface over which
the absorbed energy is re-emitted, i.e.\ $f=1/4$ if the thermal energy is emitted over
the entire $4\pi$ of the planet.  Of course, planets may be self-luminous as well,
particularly if they are young and have retained significant residual heat from formation.

The form for $\Phi_{{\rm ref},\lambda}$  depends on the scattering
properties of the planetary atmosphere.  For the case of a Lambert sphere that scatters
all incident radiation equally in all directions,
\beq
\Phi_{{\rm Lambert},\lambda}=\frac{1}{\pi}[\sin\alpha + (\pi-\alpha)\cos\alpha].
\label{eqn:lambert}
\eeq
Also for a Lambert sphere, $A_B=1$ and $A_g=2/3$.  The form for $\Phi_{{\rm therm},\lambda}$ depends on the
surface brightness distribution of the planet, which in turn
depends on the amount of heat redistribution. For the case of a tidally-locked
planet in which the absorbed radiation is promptly and locally re-emitted,
the phase curve has the same form as for a Lambert sphere \citep{seager2010}.

Resolved emission of an planet/star system is essentially equivalent
to a visual binary.  Once the overall scale of the system has been
set, measurements of the position of the planet relative to the star
at a sufficient number of epochs yield all of the orbital elements of
the system, up to the 2-fold degeneracy in orientation with respect to
the sky discussed previously. The scale of the system can be set
either by an estimate of the distance to the system, or by an external
estimate of the primary mass $M_*$ (under the assumption $M_p \ll
M_*$).  For reflected light measurements, only the product of the
geometric albedo and planet cross section can be determined;
estimating the planet radius independently generally requires an
assumption about the albedo.  For thermal light measurements, the temperature $T_p$
can (in principle) be estimated from the flux at multiple wavelengths,
and then the surface brightness can be estimated from $T_p$, and thus the radius
can be inferred from the planetary flux.  The planet mass cannot be
determined from the planet flux or its relative orbit, and must be
inferred indirectly through coupled atmosphere/evolutionary models.
In some favorable cases, mutual gravitational perturbations in
multiplanet systems may allow the determination of the planet masses
directly.

Of course, the real power of direct imaging lies in the ability to
acquire spectra of the planets once they are discovered, and thus
characterize the constituents of the planetary atmosphere.  This provides
one of the only feasible routes to assessing the habitability of 
terrestrial planets in the Habitable Zones \citep{Kasting93} of the parent stars,
and likely the {\it only} feasible route to do so for Earthlike
planets orbiting solar-type stars.

\subsection{Transits\label{sec:transits}}

The presence of a planetary companion to a star gives rise to a
multitude of physical phenomena that manifest themselves via temporal
variations of the flux of the system relative to that of an otherwise
identical isolated star.  Typically the largest of these occurs if a
fortunate alignment allows a planet to transit (pass in front of) its
host star from our perspective. In this case, the star will exhibit
brief, periodic dimmings which signal the presence of the planet.
Transits offer a intriguingly simple way to detect planets.

The condition for a transit is roughly that the projected separation
between the planet and host star at the time of inferior conjunction
of the planet is less than the sum of the radii of planet and star,
i.e., $r(t_c)\cos i\le R_*+R_p$, where $r(t_c)$ is the separation of the
planet from the host star at conjunction, and $R_*$ and $R_p$ are the
radii of the star and planet, respectively.  Given the definition of
$\omega_*$, $r(t_c)=a(1-e^2)/(1+e\sin \omega_*)$, and so transits occur
when the impact parameter of the planet's orbit with respect to the
star in units of the host radius, 
\beq 
b\equiv \frac{a\cos i}{R}\frac{1-e^2}{1+e\sin \omega_*},
\label{eqn:bimpact}
\eeq 
is less than the sum of the (normalized) radii, $b\le 1+k$,
where $k\equiv R_p/R_*$  Integrating
over $i$ assuming isotropic orbits and thus a uniform distribution of
$\cos i$ yields the {\it transit probability}, 
\beq 
P_{{\rm tr}} \equiv \left(\frac{R_*+R_p}{a}\right)\frac{1+e\sin \omega_*}{1-e^2}.
\label{eqn:ptr}
\end{equation}
For a circular orbit and assuming $k\ll 1$, this reduces
to the simple expression $P_{{\rm tr}} = R_*/a$.   Note that in these expressions,
we have used the longitude of the periastron of the orbit of star rather than the
(perhaps more intuitive) value for the planet, because the former is generally adopted
for fits to the stellar reflex radial velocity data.

When the planet transits in front of its parent star, the flux of the star
will decrease by an amount that is proportional to the ratio of the areas
of the planet and star.  For the
purposes of exposition, in the following we will assume a circular orbit, uniform
host surface brightness, and $R_p\ll R_*\ll a$ and $M_p \ll M_*$.  In the general case of a limb-darkened star, eccentric orbit,
and arbitrary scales for $R_p$, $R_*$ and $a$, the expressions for the shape of the transit
are considerably more complicated, as are the arguments for the kinds of information that can
be extracted from transit and RV signals (see \citealt{winn2010} and references therein).  However,
the basic structure of the problem is the same under our approximations, 
and what follows serves to illustrate the essential concepts.

Under these assumptions, the planet follows
a rectilinear trajectory across the face of the star with an impact parameter $b$,
and the transit signature will have
an approximately trapezoidal shape, which can be characterized by the duration $T$, 
ingress/egress time $\tau$, and fractional flux depth $\delta$. 
The depth of the transit relative to the out-of-transit flux is 
\beq
\delta = k^2.
\label{eqn:depth}
\eeq 
The duration of the transit can be quantified by its full-width at half-maximum, 
which is roughly the time interval $T$ between the two points where the center of the
planet appears to touch the edges of the star.  This is approximately,
\beq
T \simeq T_{{\rm eq}} (1-b^2)^{1/2},
\label{eqn:bigT}
\eeq
where is useful to define the equatorial crossing time (i.e, the transit duration for $b=0$),
\beq
T_{{\rm eq}} \equiv \frac{R_*P}{\pi a} = f_{{\rm tr}} P \simeq \left(\frac{3P}{\pi^2 G \rho_*}\right)^{1/3}.
\label{eqn:T0}
\eeq
Here $\rho_*$ is the mean density of the host star, and $f_{{\rm tr}} \equiv P/\pi a = P_{{\rm tr}}/\pi$ is the transit {\it duty cycle}, or the fraction of planet orbit in transit.  
The last equality, which assumes $M_p\ll M_*$, also
implies that, to an order of magnitude, the equatorial transit duration is the cube root of the product
of the orbital period and the
stellar dynamical or free-fall time ($t_{\rm dyn}\sim (G\rho_*)^{-1/2}$) squared.  

The ingress/egress time (these are equal for a circular orbit) $\tau$ is the time between
when the edge of the planet just appears to touch the star for the
first and second time (ingress, or the time between first and second ``contact'') or  
third and fourth time (egress, the time between third and fourth ``contact''), and is given by,
\beq
\tau \sim T_{{\rm eq}} \delta ^{1/2} (1-b^2)^{-1/2}.
\label{eqn:tau}
\end{equation}

One of the most useful aspects of transiting planets is that, when combined
with radial velocity data, they allow one to infer
the masses and radii of the star and planet up to a one-parameter degeneracy, as follows.
Measuring $T$, $t$, and $\delta$ from a single transit allows one to
infer $b$, $T_{\rm eq}$,
and $k$
\beq
b^2 = 1-\delta^{1/2}\frac{T}{\tau},\qquad T_{{\rm eq}}^2 = \frac{T\tau}{\delta^{1/2}}, \qquad k=\delta^{1/2},
\label{eqn:meas1}
\eeq
The impact parameter is related to the orbital inclination $i$ via
$b=a\cos i/R_*$, but $a$ and $R_*$ cannot be determined from light
curves alone.  With the detection of multiple
transits, one can further infer the period $P$, and thus the stellar
density $\rho_*$ via Equation \ref{eqn:T0}.  As reviewed in \S
\ref{sec:rv}, the reflex radial velocity orbit of the star allows one
to infer $K$ and $P$ which can be combined to determine the mass
function, $\sim (M_*\sin i)^3/M_*^{2/3}$, but a determination of the planet
mass requires both a measurement of $i$ and $M_*$.  Thus one additional
parameter is needed to break the degeneracy and set the overall scale
of the system.  This can be accomplished by imposing external
constraints on the properties of the primary, either through
parameters measured from high-resolution spectroscopy or parallax, or
invoking theoretical relations between the mass and radius of the star
through isochrones, or both.  For illustration, if we assume the primary mass
is precisely known, then we can infer $R_*$ through $\rho_*$, and $a$ through
$P$, and thus determine $i$ from the impact parameter measurement. 
Finally, we can measure $R_p$ from $k$, and the planet mass
from the mass function, $i$, and $M_*$.

\subsection{Gravitational Microlensing}

The gravitational microlensing method detects planets via the direct
gravitational perturbation of a background source of light by a
foreground planet.  When a foreground compact object (either a star or
stellar remnant) happens to pass very close to our line-of-sight to a more
distant star, the light from the background star will be split into
two images.  These images are typically unresolved, but they are
magnified by an amount that depends on the angular separation between
the lens and source.  Since this separation is a function of time, the
background source exhibits a smooth, symmetric time-variable
magnification: a microlensing event.  If the foreground lens happens
to have a planetary companion and the planetary companion happens to
have a projected separation from the primary lens near the paths of
the two primary images, the gravity of the planet will further perturb
the light, resulting in a short-lived perturbation from the primary
microlensing event, revealing the planetary companion.  Free floating
planets and planets widely separated from their parent star can also
be detected as isolated, short timescale microlensing events.

Consider a planet/star system acting as a lens located at a distance $d$ and source
located at a distance $d_s$.  Light from the source is deflected, split into multiple
images, and magnified by the gravity of the foreground
lens.  The fundamental equation that is used to derive the observable properties of 
a gravitational microlensing event is the {\it lens equation}, which relates the angular separation 
$\mbox{\boldmath$\beta$}$ between the lens and source in the absence
of lensing to the angular positions $\mbox{\boldmath$\theta$}$ of the images 
of the source created due to lensing.  For a general lens system, these are vector quantities,
but for a single lens the lens, source, and image positions are all co-linear, so we can drop
the vector notation.  The lens equation for an isolated point lens is \citep{einstein1936},
\beq
\beta = \theta - \frac{4GM_*}{c^2\drel\theta},
\label{eqn:singlec}
\eeq
where $\drel^{-1} \equiv d^{-1}-d_s^{-1}$.  If the lens and source are perfectly aligned
($\beta=0$), the source is imaged into a ring of radius equal to,
\beq
\thetae \equiv \left( \frac{4GM_*}{\drel c^2}\right)^{1/2} \simeq 713~\mu{\rm as}~\left(\frac{M}{0.5 M_\odot}\right)^{1/2}
\left(\frac{\drel}{8~{\rm kpc}}\right)^{-1/2}.
\label{eqn:thetae2} 
\eeq
The Einstein ring radius is the fundamental scale of gravitational microlensing, and depends
on the distances to the lens and source, and the mass of the lens.  
At the distance of the lens, the linear Einstein ring radius is
\beq
r_{\rm E} \equiv \theta_{\rm E}d \simeq 2.85{\rm AU}\left(\frac{M_*}{0.5 M_\odot}\right)^{1/2}
\left(\frac{d_s}{8~{\rm kpc}}\right)^{1/2}
\left[\frac{x(1-x)}{0.25}\right]^{1/2},
\label{eqn:requant}
\eeq
where $x\equiv d/d_s$.  

Normalizing by $\thetae$, the 
lens Equation \ref{eqn:singlec} simplifies to, 
\beq
u = y - y^{-1},
\label{sec:simpleplenseq}
\eeq
where $u \equiv \beta/\thetae$ and $y \equiv \theta/\thetae$.  If $u \ne 1$, this has two solutions,
$y_{\pm}=\pm \frac{1}{2}(\sqrt{u^2+4}\pm u)$, and thus in general an isolated point
lens creates two images.  One of these images is always separated by more than $\thetae$ from the lens ($y_+ \ge 1$),
and the other is always separated by less than $\thetae$ ($|y_-| \le 1$).  The separation between
the two images is $\sim 2\thetae$ and thus they are typically unresolved.  Because the images are distorted relative
to the source, they are also (de-)magnified.  The total magnification for the sum of the two unresolved images
is,
\begin{equation}
A(u)= \frac{u^2+2}{u\sqrt{u^2+4}}
\label{eqn:atot1}.
\end{equation}
The magnification increases for decreasing $u$ (better source-lens alignment), and formally diverges as $u \rightarrow 0$
for a point source.  

The source, lens, and observer are all in relative motion, and thus the angular separation
between the source and lens is a function of time: a microlensing event.  
If we approximate the relative proper motion $\mu_{\rm rel}$ of the lens and source as constant,
then we can parametrize the trajectory of the source relative to the lens as,
\beq
u(t)= \left[ \left(\frac{t-t_0}{t_\e}\right)^2+u_0^2\right]^{1/2},
\label{eqn:uoft}
\eeq
where $u_0$ is the dimensionless angular separation at the time of closest approach to the lens (the impact parameter), 
$t_0$ is the time when $u=u_0$ (also the time of maximum
magnification for a point lens), and $t_\e$ is the Einstein ring crossing time,
\begin{equation}
t_\e \equiv \frac{\theta_\e}{\mu_{\rm rel}}.
\label{eqn:tedef}
\end{equation}

\begin{figure*}
\plotone{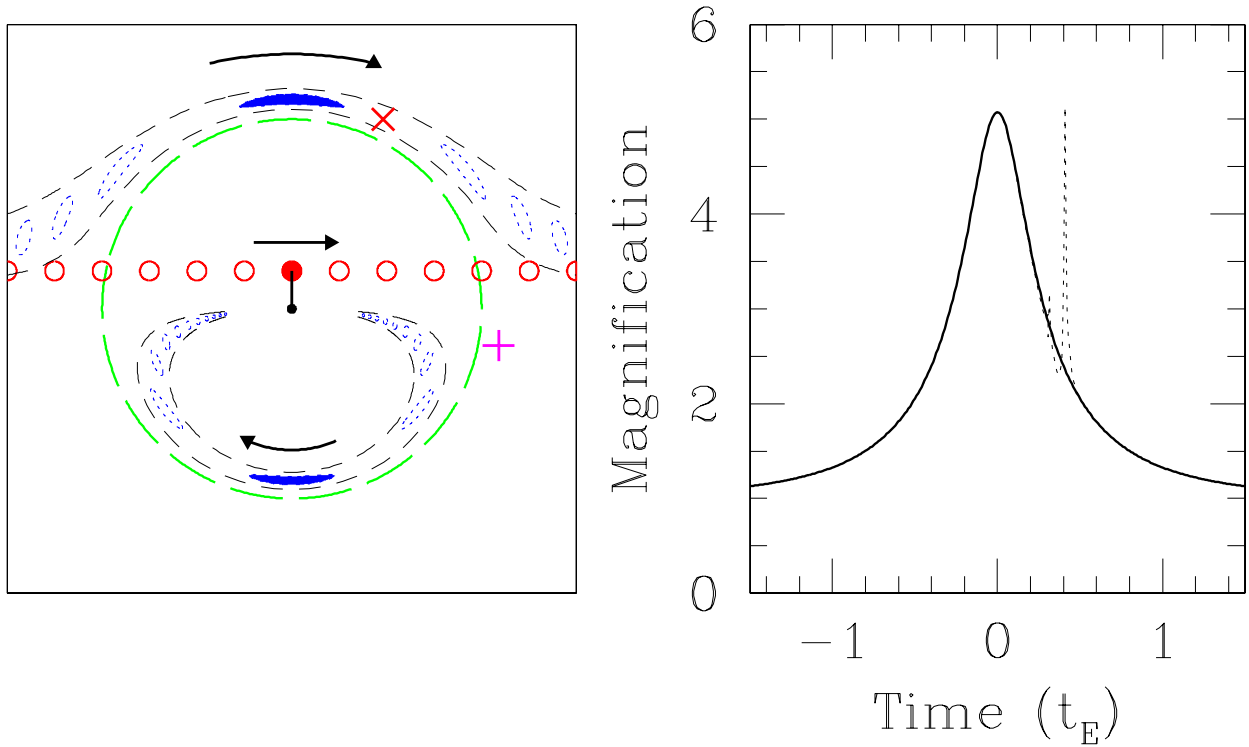}
\caption{\small 
The left panel shows the images (dotted ovals) for several different
positions of the source (solid circles) for a microlensing event with
an impact parameter of 0.2 Einstein ring radii. The primary lens is indicated as a small black dot, and the
primary lens Einstein ring is indicated a green long-dashed circle.  If
the primary lens happens to have a planet near the path of one of the
images (roughly within the short-dashed lines), then the planet will
perturb the light from the source, creating a deviation to the single
lens light curve.  Right: The magnification as a function of time is
shown for the case of a single lens (solid) and accompanying planet
(dotted) located at the position of the X in the left panel.  If the
planet was located at the + instead, then there would be no detectable
perturbation, and the resulting light curve would be essentially
identical to the solid curve.  
}\label{fig:cartoon}
\end{figure*}

Figure \ref{fig:cartoon} shows the source positions, image positions,
and magnification of an example single lens microlensing event with
$u_0=0.2$.  In general, the magnification as a function of time for a
single lens event has a smooth, symmetric form that is described by
three parameters ($t_\e, t_0, u_0$).  Events with lower $u_0$ lead to
more distorted images and higher magnification near peak.  For $u_0
\ll 1$, the peak magnification is $A_{\rm max} \propto u_0^{-1}$.
Events with $A_{\rm max} \ga 100$ are typically referred to as ``high
magnification events.''

Planetary companions to the lens star can be detected in a
microlensing event if they happen to have a projected separation in
the paths of one or both of the images created by the primary lens.  As the
image sweeps by the planet, the gravity of the planet will further
perturb the light from the source associated with the image, creating
a short-lived deviation from the single-lens form \citep{mao1991,gl1992}.  

Unfortunately, there are no simple analytic expressions relating the
observable features of planetary perturbations to the underlying
physical parameters of the planet and host star.  Adding another body
to the lens system increases the complexity of the lensing behavior
significantly, and in particular inverting the lens equation for a
binary lens to obtain the image positions for a given source position
cannot be done analytically. Furthermore, the binary gravitational
lens has a rich and complex phenomenology, which we will not attempt
to explore in this brief review.  We refer the reader to more
comprehensive summaries by \citep{bennett2008,gaudi2012}.
Here we will simply provide a qualitative discussion of
planet detection with microlensing.

Three additional parameters are required to uniquely specify the light
curve due to a binary lens (of which star/planet lenses are a
subset). The planet/star mass ratio $q=M_p/M_*$ and instantaneous
projected separation $s=r_\perp/r_\e$ between the planet and star in units
of $r_\e$ at the time of the event together specify the
magnification structure of the lens, i.e., the magnification as a
function of the (vector) source position $\mbox{\boldmath$u$}\equiv
\mbox{\boldmath$\beta$}/\thetae$.  Finally, the parameter $\alpha$
(not to be confused with the phase angle) describes the orientation of
the source trajectory relative to the projected planet/star axis.  Thus a total
of six parameters ($t_\e, t_0, u_0, q, s, \alpha$) describe the
magnification as a function of time for a binary lens and are thus
generically observable.

Single lens microlensing events yield only one parameter that depends
on the physical properties of the lens star, namely the time scale
$t_\e$. The time scale provides only a weak constraint on the lens mass,
because it depends not only on the mass, but also on the lens and
source distances, and the relative lens-source proper motion, all of
which are relatively broadly distributed for a typical
microlensing survey.  In addition, the lens 
stars are typically quite faint and are blended with other stars
(including the lensed source).  Thus little is generally known about
the host star properties.  Planetary microlensing events generally
yield two parameters that are related to the planet properties, $q$
and $s$.  While $q$ is of interest in its own right, $s$ is generally
not, because it depends on the phase, orientation, and eccentricity of
the orbit, as well as on the Einstein ring radius, all of which are a
priori unknown.  Therefore, $s$ is only weakly correlated with the
semimajor axis of the orbit, and provides essentially no constraint
on the other orbital elements.   

Although this ``baseline'' situation sounds quite dire, in fact it has
been shown that with additional effort, it is possible to obtain
substantially more information about the host star, planet, and its
orbit for the majority of detected systems using a combination of
subtle, higher-order effects that are detectable in precise
microlensing light curves, and follow-up high-resolution imaging in
order to isolate the light from the lens \citep{bennett2007}.

\subsection{Timing}

A star or stellar remnant that exhibits regular, periodic photometric
variability, such as pulsars, pulsating white dwarfs, eclipsing binary stars,
pulsating hot subdwarfs, or even stars with transiting planets, can
show evidence of a planetary companion through timing variations in
those periodic phenomena.  There are three principle sources of such
variations:  the Doppler shift, light travel time, and gravitational
perturbations.

The first of these sources is exactly analogous to the radial velocity
method, but measures changes in frequencies of some property other
than photons.  If the period of the pulsations or eclipses can be
measured to sufficient precision, then the interpretation of those
variations is identical to that in the radial velocity method.  

The light travel time effect comes about when the reflex orbit of a
star about the center of mass of the star-planet system is
sufficiently large that the additional light travel time across this
orbit is detectable as a timing variation.  This is not a truly
distinct phenomenon from the Doppler shift timing method, since it is
essentially the accumulated effects of the Doppler-shifted period that
produce the timing anomaly.  Depending on the period of the intrinsic
variation and the physical size of the star's reflex orbit, either
effect, or both, may be detectable.  

The above methods have been most successfully employed with pulsars
(through the pulse arrival times) and eclipsing binary systems
(through the timing of eclipse ingress and egress), and was
responsible for the detection of the first exoplanets \citep{Wolszczan92}.

Finally, in the case of an eclipsing system, such as an eclipsing
binary or a transiting planet, additional bodies in the system will
perturb the orbits of the eclipsing bodies.  These perturbations can
be especially large if the perturbing body is near a mean motion
resonance with the other bodies.  When applied to systems of
transiting planets this method is called {\it transit timing
  variations} (TTVs, \citealt{agol2005,TTVs,holman2005}), and has been most successfully
  employed by {\it Kepler} (e.g., \citealt{ford2012}).

\section{The Magnitude of the Problem \label{sec:order}}

By almost any physical measure, planets are small in comparison to
their parent stars, and the observable phenomena that are used to
directly or indirectly detect them are likewise small. In this
section, we attempt (where possible) to provide order-of-magnitude
estimates of the precisions of the relevant observations that are
required to detect planets using various methods.  We then use these
estimates, along with additional requirements imposed by the specifics
of the detection method (i.e, the detection efficiencies), to
provide a broad outline of the practical requirements that must be met
for planet surveys to successfully detect planets with a given set of
properties.

In general, specifying the criteria needed to detect a planet 
requires a detailed analysis of the signal and data properties, 
as well as a quantitative definition of the meaning of a detection.
However, for many of the detection methods, a rough estimate can be
obtained by decomposing the primary observable signal into two conceptually different
contributions: an overall scale and detailed signal waveform.  
The overall scale, which depends on the physical parameters of the system,
encodes the order-of-magnitude of the signal and largely dictates its
detectability.  The waveform itself depends on more subtle details of
the system (i.e., the precise shape of the planet orbit), but
typically takes on values of order unity and thus has a relatively small
effect on the detectability of the signal.  Therefore, in most cases,
these two contributions can be fairly cleanly separated. In this approximation, the
detectability of a planet with a given set of properties therefore
primarily depends primarily on the overall signal scale, and the data
quality and quantity, i.e., the typical observational uncertainties
and the total number of observations.  With this in mind, given a signal
amplitude $A$, number of observations $N$, and typical measurement uncertainty
$\sigma$, the detectability will depend primarily on the total signal-to-noise ratio 
$\snr$, which scales as,
\beq
\snr \simeq g\sqrt{N}\frac{A}{\sigma}
\label{eqn:qgen}
\eeq
where $g$ is a factor of order unity that depends on the details of the signal.   

\subsection{Radial Velocities}

The differential radial velocity signal of a planet has the form
$\Delta V_r = KF(t;e,\omega_*,T_0,P)$, where $F(t)$ encodes the
detailed shape of the RV signal.  Assuming uniform and dense sampling
of the RV curve over a time span that is long compared to $P$, and
assuming a total of $N$ observations each with measurement uncertainty
$\sigma_{RV}$, the total signal-to-noise ratio is 
\beq 
(\snr)_{RV} \simeq
g(e,\omega_*) \sqrt{N}\frac{K}{\sigma_{RV}}.
\label{eqn:qrv}
\eeq 
For a circular orbit, $g = 2^{-1/2}$, and is generally a weak function of
$e$ for $e \la 0.6$.  For larger eccentricities, $g$ declines
gradually, but more importantly the stochastic effects of finite sampling
become significant for typical values of $N$ (e.g.,
\citealt{otoole2009,cumming2004}).  For planets with periods larger than
the duration of the observations, the detectability depends additionally on the period and 
phase of the planet, and generally decreases
dramatically with increasing period, typically as $(\snr)_{RV} \propto P^{-1}$ 
(e.g., \citealt{eisner2001a,cumming2004}).

Thus, a robust detection of a planet via RV typically requires
achieving radial velocity precisions of $\sigma_{RV} \ll K N^{1/2}$.  For $M_p\ll M_*$, the semiamplitude $K$ is,
\beq
K = \left(\frac{P}{2\pi G}\right)^{-1/3}\frac{M_p\sin i}{M_*^{2/3}}(1-e^2)^{-1/2}
\label{eqn:kexp}
\eeq

Thus to detect a true Jupiter analog (i.e. a Jupiter-mass planet in a 11.8 yr, circular orbit around a
Solar-mass star), for which $K\simeq (12.5 {\rm\phantom{.}m/s}) \sin i$,
requires a few dozen observations with precisions of a few m/s.
An RV precision of $3 {\rm\phantom{.}m/s}$ corresponds to a
Doppler shift of $K/c \simeq 10^{-8}$. 
The motion induced by an Earth analog is smaller by a factor of
$318/(11.8)^\frac{1}{3} \sim 100$, so requires an additional two
orders of magnitude in precision.

Typically the centroid of stellar spectral lines at fixed equivalent width
can be measured with a precision of $\propto \sigma_V^{3/2}/N_{\rm eff}^{1/2}$, where
$\sigma_V$ is the effective velocity width of the spectral line and $N_{\rm eff}$ is
the effective number of photons in the line (i.e., the equivalent
width of the line times the photon rate per unit wavelength).
Maximizing the precision requires that the lines are
well-resolved, and thus that the instrumental velocity resolution is
less than the intrinsic velocity width of the star.  For reference,
the typical width of a spectral feature in a slowly rotating star is
of order a few km/s ($\sim 10^{-5})$, and thus resolving powers of ($R
= \Delta \lambda / \lambda \sim 10^{5}$) are needed, comparable to the
resolving power of a typical high-resolution astronomical echelle spectrograph.
The velocity precision per line is generally insufficient to detect
planetary companions, and thus averaging over many lines is required.
The statistical signal-to-noise  
ratio requirements are quite stringent, and thus bright stars and/or
large apertures are generally needed.

Because the velocity precisions needed to detect 
planetary companions are well below the intrinsic widths of the spectral lines
and even below the velocity precisions that can be obtained for individual
lines, getting close to the photon limit requires excellent control
of systematics. One of the most severe requirements is that the wavelength calibration must be both more
precise than the desired velocity precisions, and stable over many times the orbital period of
the planet.  For a Jupiter analog, this wavelength calibration must be
at a level of better than $10^{-3}$ of a resolution element, and
stable over the course of decades.  Since the Earth's motion about the
Sun imparts a periodic Doppler shift of order 30 km/s ($v/c =
10^{-4}$), this accuracy and precision must be maintained even as the
spectral lines move annually by $10^4$ times the measurement
precision.

There are at least two proven\footnote{Another technique, externally
  dispersed interferometry \citep[EDI][]{ErskineTEDI2}, has shown promise as
  a third path to precise velocimetry.  It employs an interferometer
in front of a spectrograph at modest resolution, generating a known, unresolved, sinusoidal
transmission function, somewhat analogous to a gas cell's absorption
properties.  The phase of the beating of the stellar spectrum against
this pattern is a measure of radial velocity.} paths to 
surmounting this challenge:  though precise instrumental calibration with an absorption cell (the
iodine technique), and through instrumental ultra-stability (as
exemplified by HARPS), both of which are briefly described in \S\S\ref{Campbell}--\ref{Mayor}.  

\subsection{Astrometry}

The magnitude of the differential astrometric offset of a star 
at a distance $d$ due to a planetary
companion has the general form $\Delta \theta_* = \theta_{*} F(t;e,\omega_*,i,T_0,P)$,
where the semi-amplitude of the astrometric offset for a circular, face-on orbit is,
\beq
\theta_{*} \equiv \frac{a}{d}\frac{M_p}{M_*},
\label{eqn:thetas0}
\eeq
and we have assumed $M_p \ll M_*$.   Again assuming uniform and dense sampling
of the astrometric curve over a time span that is long compared to $P$, and
assuming a total of $N$ observations each with measurement uncertainty
$\sigma_{AST}$, the total signal-to-noise ratio is 
\beq 
(\snr)_{AST} \simeq
g(e,\omega_*,i) \sqrt{N}\frac{\theta_{*}}{\sigma_{AST}}.
\label{eqn:qast}
\eeq 
We note for simplicity we have assumed that each observation yields a given uncertainty $\sigma_{AST}$ 
on the magnitude of the vector position of the star relative to some reference frame; in reality
each of these measurements may require a separate measurement
for each of the two orthogonal directions. For $e=0$,
$g(i)=[0.5(1+\cos^2 i)]^{1/2}$.  For more general
cases, the behavior of $g$ is qualitatively similar
to that for radial velocity signals.  For $e \ne 0$, $g$ depends additionally
on $\omega_*$ and $e$, but is a relatively
weak function of $e$ for $e \la 0.6$.  However, the effects of finite sampling start
to become more important as $e$ increases, particularly for low $N$. 
When $P$ is greater than the span of observations, the detectability also depends 
on $T_0$ and $P$, generally decreasing rapidly with increasing period, also typically
as $P^{-1}$.

The magnitude of the astrometric signal of a Jupiter analog orbiting a
nearby solar-type star at a distance of $D \sim 20$ pc is
$\theta_{*} \simeq 0.25~\mbox{mas}$, whereas for an Earth analog the
astrometric wobble is over 1500 times smaller, or around 0.15 $\mu$as.
Thus astrometric precisions of order a mas or a $\mu$as are needed to
detect gas giants or terrestrial planets, respectively.  Since
astrometry is most sensitive to planets orbiting the nearest stars,
which have typical proper motions of $\sim 10^3$ mas/yr and annual
parallactic motion of $\sim 10^2$ mas, the target stars typically move
by more $10^{3}$ times the required measurement precision over the course of a
year, and secularly at $10^{5}$ times the measurement precision per
decade.

The photon limit of an astrometric measurement of a star depends on
the signal-to-noise ratio and width of the point spread function
(PSF), and scales as $\sigma_{AST} \sim {\rm FWHM}/\sqrt{N}$, where
$N$ is the total number of photons in the measurement.  As mentioned
previously, diffraction limited PSFs, ${\rm FWHM} \sim \lambda/D$,
where $D$ is either the aperture of the telescope or the baseline of
the interferometer.  Baselines of $\la 100$m therefore yield single
measurement precisions of $\la 4$mas.  Therefore, the astrometric
detection of planets generally relies on the ability to achieve both
nearly photon-limited performance when measuring the centroid of
individual images, and the ability to average many individual
measurements to improve the final precision.  As is the case with RV,
excellent control of systematics is therefore required.  There are a
number of ways to achieve this, depending on the nature of the
observing setup (direct imaging, interferometry, etc.).

Interferometric methods in particular allow precisions below 1 mas
from the ground around bright stars with good, nearby reference stars,
putting astrometric exoplanet detection within reach.  Much better
control of systematics is in principle possible from space, and thus
space-based interferometers should be able to achieve precisions of
1--10 $\mu$as, making them a potential route for the detection of
nearby true Earth analogs \citep{unwin2008}.

\subsection{Imaging\label{sec:oomimaging}}

The flux ratio of a planet (or planet/star contrast) at a given
wavelength $\lambda$ and epoch can be expressed as $f_\lambda = f_{0,\lambda}\Phi(\alpha)$,
where $\Phi(\alpha)$ describes the phase curve, whose form depends on 
the properties of the planet atmosphere, and is a function of the phase
angle $\alpha$, which in turn depends on the measurement epoch and orbital
elements $e,\omega_p,i,T_0,P$.  The phase curve typically takes on values 
$\la 1$, and thus the magnitude of the reflected light signal is 
characterized by $f_{0,\lambda}$.
This factor depends on the nature of the planetary emission, but for reflected light and
thermal emission takes the form (see equations \ref{eqn:frref} and \ref{eqn:frthermal}),
\beq
f_{0,\lambda} = A_{g,\lambda} \left(\frac{R_p}{a}\right)^2 \qquad ({\rm Reflected}), \qquad
f_{0,\lambda} \simeq \left(\frac{R_p}{R_*}\right)^2 \frac{T_p}{T_*} \qquad ({\rm Thermal}),
\label{eqn:f0}
\eeq
where the latter equality assumes observations on the Rayleigh-Jeans tail, which 
yields the largest flux ratio. 
Further, for thermal emission arising from reprocessed starlight, 
\beq
f_{0,\lambda} \simeq \left(\frac{R_p}{R_*}\right)^2 \left(\frac{R_*}{a}\right)^{1/2} [f(1-A_B)]^{1/4} \qquad ({\rm Thermal, Equilibrium})
\label{eqn:f0repro}
\eeq

The signal-to-noise ratio with which a planet can be directly detected in $N$ measurements
is \citep{kasdin2003,brown2005,agol2007},
\beq
(\snr)_{dir} \simeq g \sqrt{N} \frac{f_{0,\lambda}}{\sigma_{\rm eff}}.
\label{eqn:qdir}
\eeq
Here $g=\left[N^{-1}\sum_k \Phi(\alpha_k)^2\right]^{1/2}$ is the root-mean-square of the phase function values at
the times $k$ of the observations, and $\sigma_{\rm eff}$ is the average effective per-measurement
photon noise uncertainty normalized to the total stellar flux.  In the usual background-limited
case, the primary contributions to the uncertainty are residual 
light from the stellar point spread function, and local and exo-zodiacal light. In the case where the scattered light
from the star is dominant, $\sigma_{\rm eff} \sim \sqrt{C/N_*}$ \citep{kasdin2003}, where $C$ is the contrast
ratio between the intensity of the scattered light from the star in the wings of the point spread
function relative to the peak, and $N_*$ is the total number of photons collected from the
star in the measurement.  

In contrast to many radial velocity and astrometric surveys, direct
imaging surveys are generally designed with the requirement that such
that the target signal-to-noise ratio {\it per measurement} is $\ga 1$
\citep{kasdin2003}, and thus $N\sim 1$.  Achieving a sufficient \snr\ per
measurement then typically translates into a requirement that $C\la
f_{0,\lambda}$, i.e., the flux from the planet within a given aperture is larger
than the local background from the stellar PSF in the same aperture.

Young ($<1$ Gyr old), self-luminous planets can still be quite warm
(1000--2000 K), even at arbitrarily large separations from their
parent stars, making them in some sense the easiest targets for direct
imaging surveys.  For these temperatures and roughly Jupiter radii,
the planet/star flux ratios are $f_{0,\lambda} \sim
10^{-4}$--$10^{-6}$ at near-IR wavelengths, or $\Delta m \sim$ 10--15
magnitudes.  Purely in terms of overall brightness, young exoplanets
are rather easily detectable with large telescopes at infrared
wavelengths; the primary difficulty therefore lies in suppressing the
residual starlight at the position of the planet in order to achieve the contrast ratios $C$ needed
to distinguishing the planetary light from the star's.

Since the albedos of exoplanets at distances of $\ga 0.1$ AU are
typically expected to be of order unity, the flux ratio of a planet in
reflected starlight is $f_{0,\lambda} \sim (R_p/a)^2$.  For a Jupiter
analog this is $\sim 10^{-8}$ or about 20 magnitudes, whereas for an Earth
analog this is $\sim 10^{-9}$, or about 23 magnitudes.  The bolometric
thermal flux ratio of an exoplanet in equilibrium with the
starlight will be of the same order of magnitude as the reflected
light flux ratio, however the monochromatic thermal flux ratio may be
substantially larger, since the planet is cooler and so its thermal
emission will peak in the
Rayleigh-Jeans tail of the stellar blackbody emission.  For a Jupiter
analog at $\sim 10 \mu{\rm m}$, $f_{0,\lambda} \sim 10^{-8}$, where as for
a Earth analog it is $f_{0,\lambda} \sim 10^{-7} $.

Achieving a given contrast ratio is generally more difficult for 
small angular separations from the host star, and becomes generally 
impossible closer than some minimum {\it inner working angle}, $\theta_{{\rm IWA}}$.  
Thus the angular separation $\Delta \theta=r_\perp/d$ is another important parameter
that determines the detectability of planets by direct imaging.  The probability
distribution of the projected separation $r_\perp$ given a value of $a$
for random orbital phases and viewing geometries is generally sharply peaked
at $a$.
Thus the typical angular separation of a planet with $a=5.2$ AU orbiting
a star at 20 pc is $\sim a/d = 250$ mas, whereas it is $50$ mas for
$a=1$ AU.    The inner working angle typically scales as (and is generally similar
to) the diffraction limit of telescope, 
\begin{equation}
\theta_{\rm diff} \sim \lambda / D
\end{equation}
where $D$ is the diameter of the telescope (or the most widely
separated components of an interferometer).  This corresponds to 60
mas at 2 microns on an 8m telescope.  Thus, surveys for Earth analogs
are generally only feasible for the nearest stars.

The detectability of a planet by direct imaging depends on a
complicated interplay between many variables, including the semimajor
axis and size of the planet, age and distance to the star, and
wavelength capabilities of the imaging system.  For
example, for reflected light surveys, the orbital separation effects
the detectability of the planet through the opposing effects of
contrast and angular separation.  As another example, while younger
planets tend to be more luminous, younger stars are also less common
and so typically more distant.  Additional factors may also contribute
to these interplays, such as the brightness of the exozodiacal light
as a function of semimajor axis, and variations in the planetary atmospheric
properties (e.g., albedo and absorption bands) as a function of 
semimajor axis, age, and surface gravity.  Direct imaging surveys
therefore need to be designed carefully in order to maximize the
discovery space and so chance of success.  Combined with the technical
challenges associated with achieving the contrast ratios and inner
working angles needed for planet detection briefly described below, it
is clear that direct imaging is a generally expensive and challenging
detection method.  Nevertheless, the potential payoff is enormous,
particularly when considering the goal of directly imaging Earth
analogs.

The technical aspects of imaging exoplanets comprise surmounting three challenges:
corralling starlight into a nearly diffraction-limited PSF (and away
from the planet image); mechanically blocking the starlight before it
can diffract into the planet image; and subtracting the remaining starlight at
the position of the planet image on the detector to reveal the planet
image beneath. These three challenges are most
forcefully attacked using adaptive optics, coronagraphy, and various forms
differential imaging, respectively.     

Adaptive optics (AO) refers to controlling the wavefronts of the incoming starlight
and planet-light, which ideally consist of parallel planes
propagating toward the telescope.  The atmosphere and telescope optics
both introduce aberrations to this 
wavefront which result in a PSF that differs significantly from that
which a theoretically perfect optical system would produce (for an unocculted
circular aperture, this would be an Airy function).  For most
ground-based telescopes, the primary source of wavefront 
aberrations is the atmosphere.  Adaptive optics use movable or deformable
mirrors which can be rapidly actuated in response to measured
atmospheric aberrations, usually at tens to thousands of Hertz.  These
systems dramatically reduce the effects of atmospheric blurring, and
the best of them can collect most of a star's light into the shape
dictated by optical diffraction.  This heightens the peak of faint
sources and reduces noise from the star outside the diffraction limit. 

The technique of blocking the light of a bright source to reveal faint
surrounding features is called coronagraphy.  A coronagraph uses a series of masks in
an optical system to block, reorganize, or alter the phase of
incoming light such that ``on-axis'' light from the star is almost
entirely blocked or caused to destructively interfere, while
``off-axis'' light (for instance, from a nearby planet) is relatively
unaffected.  Because important aspects of this technique happen in the
pupil plane, stellar photons can be distinguished from planetary photons
and rejected before they arrive on the same pixels on the detector.  The
effect is to reduce the contamination from stellar photons at the
detector position of the planet, enhancing its detectability outside
of the diffraction limit.   There has been a proliferation of
coronagraph designs in recent years, but they share the common feature
of reducing or controlling the nature of the diffraction of the light
into the planet image.
 
Adaptive optics systems and coronagraphs are not perfect.  Their limitations, the
aberrations introduced by the telescope, and diffraction spikes and
rings from the aperture can result in significant amounts
of starlight outside of the diffraction limit.  The most insidious of
these effects are the semi-static patterns of ``speckles'' from residual
wavefront errors.  Differential imaging is the process of precisely determining
the PSF of the starlight and attempting to subtract it, leaving only
the planet light to be detected.  In principle, differential imaging
is limited by the quality of the model PSF and the unavoidable photon
noise in the residuals to that model.  The reference image being
subtracted can be determined from a reference star (RDI), or from the
data themselves through angular modulation (ADI), spectral analysis
(SDI), polarization analysis (PDI), or other some other method or
combination of methods.

A conceptual cousin of coronagraphy is interferometry, which allows
widely separated apertures to combine incoming light to form interference fringes whose
amplitudes and phases are sensitive to the presence of faint, off-axis
companions.  Such work is common at radio wavelengths, and in the
infrared can be especially profitable just inside of the traditional
diffraction limit of the telescope.   Two such techniques are aperture
masking interferometry, where a single telescope pupil is divided into
small sub-apertures and the light is combined at the focal plane, and
nulling interferometry where light from two telescopes is combined
such that the starlight undergoes destructive interference, while the
planet light, incoming at a slightly different angle, interferes
constructively.  

\subsection{Transits}

The fractional change in the flux of a star when it is transited by a
planetary companion has the form $\Delta F_*/F_* = -\delta\,\,
F(t;R_p,M_*,R_*,a,i,e,\omega_p)$, where $\delta=k^2$ is the square of
the planet/star radius ratio, and the function $F(t)$ describes the
detailed shape of the transit curve, and also generally depends on the
surface brightness profile of the star.  In the case of circular
orbits, no limb darkening, and $R_p \ll R_*$, the form for $F(t)$ can be approximated by a
box car with a depth of unity and duration of $T=T_{eq}(1-b^2)^{1/2}$,
where $T_{eq}$ and $b$ are the equatorial crossing time and impact
parameter, respectively, as defined in \S \ref{sec:transits}.  The fraction of time the
planet is in transit (the transit duty cycle) is then $f_{tr} = T/P$.

Transit surveys generally operate by obtaining many observations of
the target stars over a given time span.  Of course, in order to be
detectable a planet must be favorably aligned such that it transits.
The transit probability is roughly $P_{tr}\sim R_*/a$.  Then, assuming
uniform sampling over a time span that is long compared to $P$, the
signal-to-noise ratio of the transit when folded about the correct
planet period is,
\beq
(\snr)_{tr}\simeq \sqrt{N f_{tr} }\frac{\delta}{\sigma_{ph}} 
\label{eqn:qtr}
\eeq
where $N$ is the total number of observations and $\sigma_{ph}$ is the 
fractional photometric uncertainty.    Therefore, the probability of detecting
a given planet via transits can be roughly quantified by three characteristics of the planetary system
that depend primarily on $R_p, R_*$, and $a$,
\beq
\delta = \left(\frac{R_p}{R_*}\right)^2, \qquad P_{tr} \sim \frac{R_*}{a},\qquad f_{tr} \sim \frac{P_{tr}}{\pi},
\label{eqn:trandetspars}
\eeq

For a typical hot Jupiter with $R_p \simeq R_J$ and $P\simeq 3$ days
orbiting a solar-type star, the transit probability is $P_{tr} \sim
10\%$, the transit depth is $\delta \sim 1\%$, and the duty cycle is
$f_{tr} \sim 3\%$.  These parameters place Hot Jupiters well within
the capabilities of ground-based surveys, although the requirements
are not trivial.  First, since Hot Jupiters are only found around
$\sim 0.5\%$ of solar-type stars \citep{gould2006}, many thousands of stars must
be surveyed to guarantee a transiting Hot Jupiter.  Obtaining relative
photometry at precisions of less than a few millimagnitudes for
thousands of stars simultaneously from the ground is generally
difficult, and thus ground-based transit surveys operate close to the
limit where $\delta/\sigma_{ph} \sim 1$.  Therefore, hundreds of
epochs during transit are needed for robust detections, corresponding
to many thousands of total measurements.  Aliasing effects arising
from the diurnal constraints make achieving the required number of
points in transit more challenging.  All of these requirements are
most easily met with relatively small aperture, but very wide
field-of-view telescopes (e.g., \citealt{pepper2003,bakos2004,pollacco2006,mccullough2005}).

In fact, finding transiting planets in wide-field surveys has proven
even more difficult than simply meeting these (already difficult)
requirements.  First, wide-field transit surveys must contend with a
huge fraction of false positives in the form of grazing eclipsing
binaries (EBs), eclipsing binaries blended with brighter stars, and
more exotic variables.  Furthermore, even those signals that are
consistent with a Jupiter-sized transiting object can, in principle,
be much more massive companions, since the
radius of compact objects is essentially constant from the mass of
Saturn through $\sim 0.1 \Msol$ (e.g.\ \citealt{burrows1997}).  Thus radial 
velocity follow-up is need to eliminate
these false positives.  Finally, high-precision (few m/s) radial velocity follow-up is needed
to precisely measure the planet mass. The most successful searches achieve
reliably high photometric accuracy over large fields, and employ
multiple sites with good longitudinal coverage, sophisticated and
automated transit identification algorithms, and thorough follow-up
campaigns that using multi-band photometry, multi-band astrometry (to
rule out close, chance alignments of EBs and foreground stars) , and
radial velocity work.  Further characterization of a transiting planet
is most successfully done using photometry and high signal-to-noise
ratio spectroscopy with larger ground and/or space telescopes.

The requirements for the detection of Earth analogs orbiting solar-type stars are
especially challenging.   In this case, the fractional transit depth is $\sim 10^{-4}$, the
transit probability is $\sim 0.5\%$, and the duty cycle is $\sim 0.1\%$ (i.e.,
the planet transits for $T\la 13$ hours once a year).  The detection of
transiting Earth analogs requires essentially continuous observations
of hundreds of stars, and precisions of better than $\sim 0.1$ mmag for periods
of several years.  These requirements cannot be met
from the ground, and require space-based photometric monitoring.  
Indeed, the {\it Kepler} mission was
designed to detect such planets, achieving the required photometric precision
to detect Earth-sized planets on tens of thousands
of stars \citep[][see also \S~\ref{spacetransit}]{borucki2010}.

Transiting planets can also be found via photometric follow-up of
known radial velocity companions; indeed the first transiting planet
was discovered in this way  \citep{HenryG00,Charbonneau00}.  Here the challenges
are somewhat different than the ``traditional'' method of discovering
transiting planets through their photometric signature.  First,
the probability that a given radial velocity companion will also
transit its parent star is low, $\la 10\%$.  Second, radial velocity
searches have traditionally been limited to relatively bright stars,
and so the total number of stars have targeted for precision RV
searches is $\sim 10^3$, making the total yield of transiting planets
from this sample also low.  Furthermore,
achieving photometry at the level of precision needed to detect the
transit signature may be challenging for very bright stars, primarily
because of the lack of suitable comparison stars.  Finally, the uncertainties
in the predicted times of inferior conjunction from the radial
velocity fits can be quite large, from several hours to several days. 
Nevertheless, seven transiting systems have been discovered amongst
the sample of companions first discovered via radial velocity, and there are
ongoing projects that aim to increase this sample by first refining
the radial velocity emphemerides of promising systems, and 
then performing photometric follow-up \citep{kane2009}.  

\subsection{Microlensing}

Unlike the detection methods discussed above, the signals caused by
planetary companions in microlensing events cannot be described
analytically except in a few specific limits that are not generally
applicable.  Nevertheless, we can provide some qualitative guidelines
and approximate scaling behaviors that will elucidate the general
requirements for successful surveys for planets with microlensing.
We stress that, because of the large diversity in the properties
of the systems that give rise to gravitational microlensing events, 
the expressions provided should be treated as very rough estimates only.

A somewhat unusual attribute of the microlensing method is that the
magnitude of a microlensing perturbation does not depend on the
properties of the planet in the general case.  Rather, the magnitude
depends primarily on the angular separation of the planet from the
image(s) it is perturbing.  However, the duration of the planetary
perturbation does depend on the planet properties, in particular the
mass ratio $q$.  Very approximately, the duration of the planetary
deviation is $\Delta t_p \sim q^{1/2} t_\e$, where $t_\e$ is the
primary event time scale.  The primary event light curves must be
sampled on a time scale significantly smaller than $\Delta t_p$ in
order to detect and characterize the planetary perturbation.
Furthermore, the detection probability also depends on the planet mass
ratio, such that $P_{det} \sim 20\%(q/0.001)^{\sim 5/8}$ 
\citep{horne2009}.  This detection
probability is averaged over a uniform distribution of impact
parameters, and is appropriate for planets with projected separations
that are within a factor of $\sim 2.6$ of the Einstein ring radius,
$r_\perp \sim [0.6-1.6] d \theta_\e$, sometimes called the
``lensing zone''. Planets with separations much smaller or much larger
than this range have substantially lower probability of detection.  As
discussed in the context of direct imaging, the distribution of
projected separation $r_\perp$ for random viewing geometries and
orbital phase is sharply peaked at $r_\perp \sim a$.

In addition, there is a minimum mass that can be detected in
microlensing surveys, that is set by the finite size of the source
stars.  When the angular size of the planet perturbation region is
substantially smaller than the angular size of the source, the planet
perturbs only a small fraction of the source, and the magnitude of the
resulting deviation is strongly suppressed.  A rough limit on the mass
ratio can be established by when the angular size of the source
$\theta_*$ is a factor of $\sim 3$ times larger than the angular
Einstein ring radius of the planet $\theta_p = q^{1/2}\theta_\e$, corresponding 
to roughly an order of magnitude suppression of the planet
signal.  Thus $q_{min} \sim 0.1 \rho_*^2$, where $\rho_* \equiv
\theta_*/\theta_\e$ \citep{gould1997}.

Thus the parameters that determine the detectability of planets with microlensing are
\beq
\Delta t_p \sim \left(\frac{M_p}{M_*}\right)^{1/2} t_\e,\qquad P_{det} \sim 20\% \left(\frac{M_p/M_*}{0.001}\right)^{~5/8},
\qquad a \sim [0.6-1.6] d \theta_\e, \qquad q_{min} \sim 0.1 \rho_*^2,
\label{eqn:microdet}
\eeq
where the parameters $t_\e$, and $\rho_*$ additionally depend on the
mass and distance to the host star via the angular Einstein ring
radius (see Eq.~\ref{eqn:thetae2}).  The distributions of $M_*$,
$t_\e$, $d$, and $\theta_\e$ for microlensing events toward the
Galactic bulge are quite broad, but we can take typical values of $M_*
\simeq 0.5$ $\Msol$, $t_\e \simeq 25$ days, $d \sim 4$ kpc, and
$\theta_\e \sim 0.7$ mas. Thus microlensing planet surveys are most
sensitive to planets with semimajor axes of $a \simeq [1-5]~{\rm AU}
(M/0.5~\Msol)^{1/2}$.  For a Jupiter-mass planet, the typical planet
perturbation duration is $\Delta t_p \sim 1$ day, and the typical
detection probability is $\sim 30\%$ in the lensing zone.  For an
Earth-mass planet, $\Delta t_p \sim 1.5$ hours, whereas the detection
probability in the lensing zone is $\sim 1\%$.

The typical dimensionless source size for a clump giant star ($\sim
13$ $R_\odot$) in the Galactic bulge is $\rho_* \sim 0.01$, whereas for a
turn-off star ($\sim R_\odot$) it is $\sim 0.001$.  Thus the minimum
mass ratio that can be detected by monitoring clump giant sources is
$q_{min} \sim 10^{-5}$, corresponding to $\sim 1.7\times$ mass of the Earth for a
typical primary lens of $0.5~\Msol$.  For main sequence stars in the
bulge, $q_{min} \sim 10^{-7}$, corresponding to just over the mass of the Moon!
Thus detecting planets with mass of the Earth or less requires
monitoring main-sequence stars.  The difficulty lies in the fact, in
the crowded fields toward the Galactic bulge, most main sequence stars
are severely blended with other unrelated background stars in typical
ground-based seeing conditions, dramatically increasing the
photometric noise.  Therefore, detecting planets with mass
substantially less than that of the Earth generally requires a
space-based survey \citep{bennett2002,bennett2008}.

A final difficulty with in microlensing surveys
is the low overall event rate of gravitational microlensing events.
Toward the Galactic bulge, the rate of microlensing events is roughly
$\Gamma \sim 10^{-5}$ per star per year (e.g., \citealt{kiraga1994}).  Thus, in order to detect
$\sim 10^{3}$ events per year (the current number of microlensing
events that are detected per year toward the Galactic bulge by the
Optical Gravitational Lensing Experiment (OGLE) collaboration\footnote{See
http://ogle.astrouw.edu.pl/ogle4/ews/ews.html.}), of order 100 million
source stars must be monitored. There are 3 million stars per square
degree down to an $I$ magnitude of 19 in Baade's window
\citep{holtzman1998}, where $I\sim 19$ is roughly the peak of the
distribution of baseline magnitude for microlensing events.  Thus 
several tens of square degrees of the bulge must be monitored.

The unpredictability of microlensing events requires monitoring the
potential source stars with a cadence that is substantially smaller than the
timescale of interest. For the primary microlensing events, which have
a typical $t_\e \sim 25$ days, this means roughly daily observations.
Detecting the planetary perturbations on these events requires much
higher cadences of a few hours or less.  Furthermore, since the total
durations of the planetary perturbations are of order a day or less,
networks of longitudinally-distributed telescopes must be employed in
order to avoid missing part or all of the perturbations.  Given these
requirements, traditional microlensing planet surveys have used a
two-tier approach, where collaborations with dedicated access to
telescopes with a relatively wide fields of view of $\sim 0.5-2$
square degrees monitor the tens of square degrees needed to detect the primary
microlensing events, but with cadences that are generally insufficient
to detect planetary perturbations on these events
\citep{udalski2003,sako2008}.  These survey collaborations
alert the microlensing events real-time before the peak magnification, thus allowing
``follow-up'' collaborations with access to narrow-angle telescopes on
several continents to monitor only a subset of the stars that display
ongoing microlensing events with the cadences needed to detect
planetary perturbations \citep{albrow1998,tsapras2009,dominik2010,gould2010}.
Future surveys will operate on a very different principle, as 
described in \S \ref{sec:futuremicro}.

\subsection{Timing}

The magnitude of the signal in other planet detection techniques
varies.  Timing for millisecond pulsars like PSR 1257+12 can in
principle detect extremely low mass objects (significantly $<
1\mearth$) given a sufficient amount of data, limited primarily by
pulsar timing noise.  

Other timing techniques, such as eclipsing binary times or pulsating
hot subdwarfs, rely on timing variations being correctly interpreted 
as a light-travel time effect of a star or stellar system orbiting a common
center of mass with an unseen companion.  A summary of such detections is listed in
\citet{sdBs}; most of these detections imply minimum companion masses of several
times that of Jupiter.  The sensitivities of these methods is difficult to 
determine, however, since they depend on the magnitude of all non-orbital
origins in timing variations, which have not been well quantified.  Most of the current
detections are of fewer than two full apparent orbits (periods are
3--16 yr) and so the strict periodicity that is characteristic of Keplerian signals cannot
yet be confirmed. Further, quasi-cyclical timing variations may be generated by poorly
understood internal mechanisms such as the ``Applegate effect''
\citep{Applegate92}.  Following these apparent planetary systems for
multiple orbits will help illuminate the true sensitivities of these methods.

Transit timing variations (TTVs) provide an extremely sensitive method
of detecting new planets or characterizing known planets in a
transiting system.  The sensitivity is a complex function of the
orbital parameters of the planets involved, but is optimized when the
planets are in mean-motion resonances \citep{Veras11}.  {\it Kepler} is
sufficiently precise in its timing to measure variations of order
minutes in the ingress and egress times of transiting planets, which
in principle allows it to reach mass precisions of order 1 \mearth\
over several years of observation.  

In known multi-transit systems, these variations can be used to infer the
masses of the planets involved \citep[e.g.][]{Kepler-11} and in
apparently single systems they can be used to detect non-transiting
planets \citep[e.g.][]{Kepler-19}.  Ground-based planet transit timing will
generally be limited to precisions of a tens of minutes, and so have
correspondingly weaker sensitivities.

\section{Comparisons of the Methods}

In the previous sections, we reviewed the primary exoplanet detection
methods in some detail, outlining the principles of each method,
including the primary physical observables and practical challenges
associated with achieving robust detections.  In this section, we
place these discussions in the context of the larger goal of
constraining exoplanet demographics by outlining the regions of planet
and host star parameter space where each method is most sensitive.
When then compare the methods with one another in this context, highlighting
the strong complementarity of the methods.  We also briefly
discuss and compare how the intrinsic sensitivity of each method to
planets in the Habitable Zones of their parent stars scales with 
host star mass.

The sensitivity of the various detection methods as a function
of planet mass and separation is illustrated in Figure \ref{fig:exoplanets}.
We show the masses and semimajor axes of the exoplanets
discovered by radial velocities, direct imaging, timing, transits, and
microlensing, as of Dec.\ 2011.  In addition, we show estimates
of the sensitivity of various surveys using radial velocities, direct imaging,  transits,
microlensing, and astrometry.  In the following subsection, we explain the scaling 
of these survey sensitivities with planet parameters, and explain our specific
choice for their normalization.
Host star mass is a third parameter that can
strongly influence the sensitivity of these methods, but is suppressed in this figure. 
Therefore, in order to provide a somewhat fairer comparison across the broad range of host
star masses represented in this figure, we normalize the semimajor
axis by an estimate of the snow line distance (e.g., \citealt{kennedy2008}),
\beq
a_{sl}= 2.7~{\rm AU}\frac{M_*}{\Msol}.
\label{eqn:asl}
\eeq
The snow line is the location in the protoplanetary disk where the temperature
is below the sublimation temperature of water.  In the currently-favored paradigm
of planet formation, the snow line distance plays an important role, as the larger surface density
of solids beyond the snow line facilitates giant planet formation, whereas primarily rocky planets
are expected to form interior to the snow line
(e.g., \citealt{lissauer1987,pollack1996,ida2004,mordasini2009}).

Both the locations of the known planets and the regions of sensitivity
in Figure \ref{fig:exoplanets} serve to illustrate the complementarity
of the various detection methods.  In particular, radial velocity and
transit surveys are most sensitive to relatively short-period
exoplanets with separations interior to the snow line.  State of the
art surveys using these methods are sensitive to planets as small as
the Earth.  In contrast, direct imaging and microlensing surveys are
most sensitive to planets beyond the snow line. Current direct imaging
surveys are sensitive to very widely-separated, massive planets.
Current and near-term microlensing surveys are sensitive to planets
with masses greater than that of the Earth, whereas a space-based
microlensing survey would be sensitive to planets with mass greater
than that of Mars and separations greater than a few AU, including
free-floating planets.  The combined sensitivity of current and
proposed surveys using the methods discussed in the chapter
encompasses an extremely broad volume of parameter space, including
planets with masses greater than that of the Earth for arbitrary
separations (including free-floating planets), host stars with masses
from the bottom of main sequence to several times the mass of the sun,
and distances from the solar neighborhood to the Galactic center.
Thus the full complement of these methods can potentially provide a
nearly complete picture of the demographics of exoplanets.

\subsection{Sensitivities of the Methods\label{sec:sens}}

\begin{figure}
\plotone{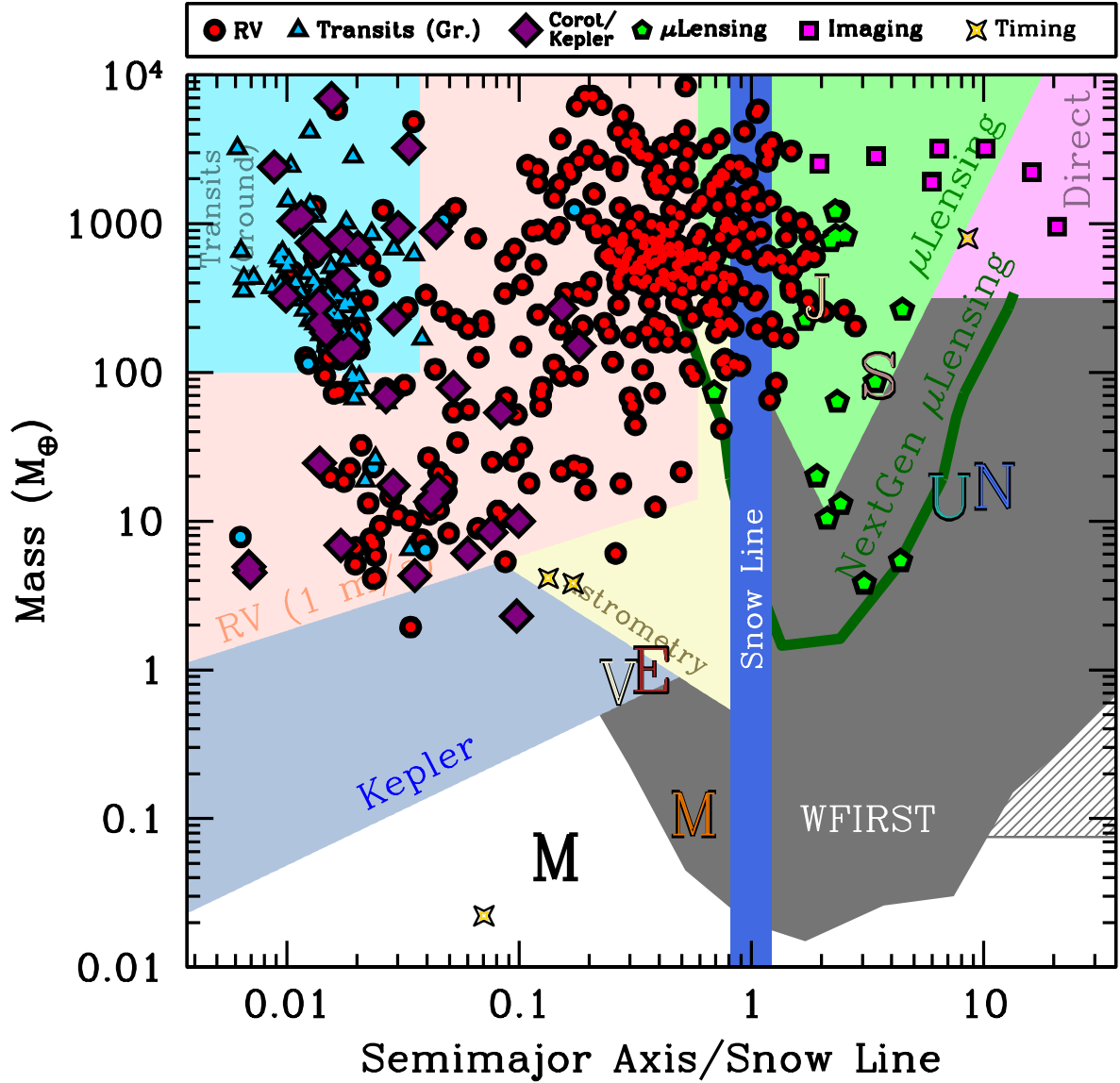}
\caption{The points show the masses versus
semimajor axis in units of the snow line distance for the exoplanets that
have been discovered by various methods as of Dec.\ 2011.  
See the Extrasolar Planets Encyclopedia
(http://exoplanet.eu/) and the Exoplanet Data Explorer (http://exoplanets.org/). Here we
have taken the snow line distance to be $a_{sl}=2.7~{\rm AU}(M_*/\Msol)$.
Radial velocity detections (here what is actually plotted is $M_p\sin i$) are indicated by red circles
(blue for those also known to be transiting), transit
detections are indicated by blue triangles if detected from the ground
and as purple diamonds if detected from space, microlensing detections are indicated
by green pentagons, direct detections are indicated by magenta squares, and detections
from pulsar timing are indicated by yellow stars.    
The letters indicate the locations of the Solar System planets. 
The shaded regions show rough estimates of the sensitivity of various
surveys using various methods, demonstrating their complementarity.
\label{fig:exoplanets}}
\end{figure}

\begin{itemize}

\item {\bf Radial Velocities}.  The radial velocity signal and signal-to-noise
ratio scales as 
\beq
(\snr)_{RV} \propto M_p P^{-1/3} M_*^{-2/3} \propto M_p a^{-1/2} M_*^{-1/2}.
\eeq
Thus, at fixed host star mass, RV surveys are generally more sensitive
to more massive planets, and have a weak preference for shorter-period
planets, at least for periods shorter than
the span of the measurements.  At fixed \snr, the minimum detectable mass scales as
$M_{p,min} \propto P^{1/3} M_*^{2/3} \propto a^{1/2} M_*^{1/2}$.  This mass limit is shown in
Figure \ref{fig:exoplanets} assuming $M_* = \Msol$ and a minimum
detectable RV signal of $K\sim 1$~m/s (i.e., corresponding to a
$(\snr)_{RV} \sim 10$ detection with 100 observations each with single
measurement precisions of $\sim 1$m/s.). In principle, sufficiently massive planets can be
detected even when they have periods longer than the duration of observations, however
it will generally not be possible to uniquely measure $M_p\sin i$
and $P$ in these cases, and thus the usefulness of such `detections' are significantly compromised.
Therefore, we simply assume an upper limit on the period of $P = 2000$ days.

At fixed planet properties,
the RV signal increases with decreasing host
star masses as $M_*^{-2/3}$.  However, there are many additional
factors that enter into the overall detectability as a function of
mass, through the radial velocity uncertainty $\sigma_{RV}$.  At fixed
distance to the host star, the velocity uncertainty due to photon
noise increases with decreasing mass for main-sequence stars, both
because of the strong mass-bolometric luminosity relation for
main-sequence stars, and because of the difficulties of continuum
normalization in very cool stars, where most RV surveys are carried out.  The intrinsic
velocity information in stellar spectra also varies as a function of
spectral type.  In particular hot ($T_{\rm eff} \ga 6500$K) stars have
few spectral lines and typically rotate much more rapidly than cooler
stars with convective envelopes.  Finally, the astrophysical radial velocity
noise (i.e., ``jitter'' due to spots) also depends on spectral type and stellar age 
\citep{wright2005}. 
When taken together, these factors tend to favor late G or early K quiet main-sequence stars as the most
promising for detecting low-mass planets (e.g., \citealt{pepe2011}).

\item {\bf Astrometry}.  The astrometric signal and signal-to-noise ratio scales
as 
\beq
(\snr)_{AST} \propto M_p P^{2/3} M_*^{-2/3} d^{-1} \propto M_p a M_*^{-1} d^{-1},
\eeq
and so at fixed stellar mass, astrometric surveys are more sensitive
to massive, long-period planets.  At fixed \snr, the minimum detectable
mass scales as $M_{p,min} \propto P^{-2/3} M_*^{2/3} d \propto a^{-1} M_* d$.  This limit
is shown in Figure \ref{fig:exoplanets} assuming $M_*=\Msol$ and
$d=10$ pc, and a minimum
detectable astrometric signal $\theta_{*}\simeq 0.35~\mu{\rm as}$
(i.e., a $(\snr)_{AST} \sim 5$ detection with 200 2-D observations each
with $\sigma_{AST} \sim 1~{\mu}{\rm as}$ precision).
As with radial velocity observations, although it is possible detect the 
astrometric signal of planets
with period larger than the duration of observations, it is generally
not possible to independently measure the mass and orbital parameters
with observations that do not cover a complete orbit.  This is
particularly problematic for astrometric observations, because 
in this regime the signal of the proper motion of the star is 
partially degenerate with the astrometric signal of the planetary companion.
Astrometric surveys are therefore expected to have the most sensitivity
to planets with periods similar to the survey duration, and increasing
the survey duration has a strong effect on the survey yield.

At fixed planet properties and host star distance, the astrometric
signal increases with decreasing host star mass as $M_*^{-2/3}$.
Stellar spots are generally not expected to be an important source of
astrometric noise \citep{makarov2009}, and thus the only additional dependence of
the sensitivity of astrometric surveys on stellar mass enters through
the effects of the typical distance and photon noise uncertainty of
the host stars.  Specifically, low-mass stars are more common and thus
have a smaller typical distance, but are less luminous and thus yield
poorer astrometric precision.

\item {\bf Imaging}.

For direct detection in reflected and equilibrium thermal emission light, 
the planet/star flux ratio and thus the signal-to-noise ratio scales as
\beq
(\snr)_{dir} \propto R_p^2 a^{-2} \qquad ({\rm Reflected}),
\eeq 
\beq
(\snr)_{dir} \propto R_p^2 T_p R_*^{-2} T_*^{-1} \qquad ({\rm Thermal})
\eeq
\beq
(\snr)_{dir} \propto R_p^2 R_*^{-3/2} a^{-1/2}\qquad ({\rm Thermal, Equilibrium}),
\eeq
where the last two forms again assume observations in the Rayleigh-Jeans
tail.  The other primary requirement for direct imaging is that the 
angular separation between the planet and star is larger than inner
working angle, and thus $a \ga \theta_{{\rm IWA}} d^{-1}$.  Thus at fixed primary
properties and distance, larger and hotter planets are more readily detectable.
As discussed in \S \ref{sec:oomimaging}, the dependence of detectability on semimajor axis is not trivial:
planets with larger orbits generally have smaller flux ratios, however they must
have an angular separation greater than the inner working angle to be detectable.
Furthermore, additional effects that are likely to depend on the planet
semimajor axis may affect the detectability, such as the variation of the planetary
albedo with separation.  

At fixed planet properties, the signal-to-noise ratio with which a
planet can be detected in reflected light is largely independent of
the host star properties.  For detection in thermal emission, larger
and/or hotter host stars generally give rise to smaller flux ratios.
The other effect of host star mass enters through the dependence on
distance: less massive host stars are more numerous and so have a
smaller average distance, whereas more massive host stars are more
luminous and thus give rise to smaller photon noise uncertainties at
fixed distance.  Finally, the age of the host star plays an important 
role in the detectability of planets, particularly with current
surveys: young, self-luminous planets have flux ratios that are 
both larger than would be expected for planets whose emission is dominated
by reflected light or equilibrium thermal emission, and are independent
of their semimajor axis.  Thus relatively luminous planets with separations
well beyond the inner working angle can be found around young stars.  

Current ground-based imaging surveys are most 
sensitive to young, massive ($M_p \ga \mjupe$) self-luminous planets with semimajor axes of 
$\ga 10$~AU around the nearest stars.  Thus these surveys are sensitive to planets in
a regime of parameter space that is not currently being probed by other methods.
We illustrate the current region 
of sensitivity of imaging surveys in Figure \ref{fig:exoplanets}, assuming planets
with $M_p \ga \mjupe$ and $a \ga 10$ AU can be detected. Future
surveys (some of which will be initiated very soon) will be sensitive
to a much broader region of planet parameter space.

\item {\bf Transits}
Assuming uniformly-sampled observations over a time span $T$ that is
long compared to the planet period, the signal-to-noise ratio with
which a transiting planet can be detected scales as 
\beq (\snr)_{tr}
\propto R_p^2 P^{-1/3} M_*^{-5/3} \propto R_p^2 a^{-1/2} M_*^{-3/2},
\eeq 
where we have assumed $R_* \propto M_*$, as appropriate for stars
on the main sequence with $M\la \Msol$.  In addition, the transit
probability scales as $P_{tr} \propto P^{-2/3} M_*^{2/3} \propto
a^{-1} M_*$, and the requirement to detect at least $n$ transits sets
a strict lower limit on the period $P \le T/n$.  Thus for fixed host
star properties, the sensitivity of transit surveys is strongly
weighted toward short period, large-radius planets.  At fixed $\snr$,
the minimum detectable planet radius scales as $R_{p,min} \propto
P^{1/6} M_*^{5/6} \propto a^{1/4} M_*^{3/4}$.  For planets with a
constant density, this translates into a minimum mass of $M_{p,min}
\propto R_{p,min}^{3} \propto P^{1/2} M_*^{5/2} \propto a^{3/4}
M_*^{9/4}$. This limit is shown in Figure \ref{fig:exoplanets},
assuming a minimum $(\snr)_{tr}=8$ and a mid-latitude transit and
photon noise-limited precision for a $M_*=\Msol$, $V=12$ star from {\it Kepler}
\citep{gilliland2011}.  The relatively strong dependence of the signal-to-noise ratio
on planet radius, combined with the decreasing transit probability
with increasing planet period, generally implies that the 
yield of a transit survey is a relatively weak function of the total duration $T$.  

Main-sequence stars are clearly the best targets for transit searches. 
At fixed planet radius and period, low-mass main-sequence stars yield 
larger transit signals.  However, for photon noise limited
uncertainties, the signal-to-noise ratio also depends on the stellar
luminosity in the wavelength of the observations, and the distance to the star.
Low-mass stars are more common, and are therefore are closer
on average. These net result of these various countervailing effects
on the scaling of the sensitivity of transit surveys with host star
mass depends on the survey parameters, particularly the wavelength,
target field, and magnitude limit \citep{pepper2005}.  All-sky surveys in the visual at
fairly bright magnitudes tend to be dominated by F and G stars
\citep{pepper2003}, but surveys for fainter stars, or surveys in near-IR
wavelengths, tend to be more heavily weighted to toward low-mass stars.

\item {\bf Microlensing}

As discussed previously, microlensing surveys are less amenable to
analytic characterization of the sensitivity and scaling with planet
and host star parameters.  Nevertheless, we have the general
qualitative result that the sensitivity peaks for semimajor axes that
are similar to the Einstein ring of the host star lens, $a_{opt} \sim
2.85~{\rm AU}(M_*/0.5\Msol)^{1/2}$.  Both the detection probability at
fixed semimajor axis and range of sensitivity around this peak
increases for larger mass ratio.  The range of planet mass and
semimajor axis to which a given microlensing survey is most sensitive
depends on the details of the survey design, in particular the number
and peak magnification of the primary events that are monitored, and
the precision and cadence of the observations.  In Figure
\ref{fig:exoplanets}, we show representative estimates for the
sensitivity of a current ground-based microlensing survey based on the
analysis of \citet{gould2010}, a next-generation ground-based survey
as described in \S \ref{sec:futuremicro}, and a space-based survey similar to WFIRST as
described \citet{green2011}. Generally, current microlensing surveys
are most sensitive to planets with $M_p\ga 10~M_\oplus$, with
separations just beyond the snow line spanning a factor of a few in
semimajor axis.  Next generation ground-based surveys will lower the
sensitivity in mass by roughly an order of magnitude to $M_p \ga
M_\oplus$, and broaden the range of semimajor axis by a factor of
$\sim 2$.  A space-based survey would be sensitive to planets with
mass greater than the mass of Mars with separations greater than a few
AU, including analogs to all of the solar system planets except
Mercury.

Because microlensing does not rely on the detection of photons 
from the host star, the sensitivity of a microlensing survey to the host
star parameters enters primarily through the microlensing
event rate as a function of host star mass.  The microlensing event
rate is given by the integral of the number density of lenses along
the line of sight to the bulge, weighted by the cross section for
lensing, which depends on the host star Einstein ring radius ($\propto M_*^{1/2}$)
and its projected transverse velocity.  Figure \ref{fig:mass}
shows a theoretical estimate of the microlensing event rate toward
the Galactic bulge, as a function of the lens mass.  The sensitivity
of microlensing planet surveys 
is roughly proportional to this event rate.  Thus microlensing
surveys are sensitive to main-sequence hosts with mass from below
the hydrogen burning limit up to the turn-off of the stellar population ($\sim M_\odot$),
as well as remnant hosts.  The sensitivity to brown dwarf and main-sequence hosts 
is roughly constant per $\log{M_*}$.  

\begin{figure}
\plotone{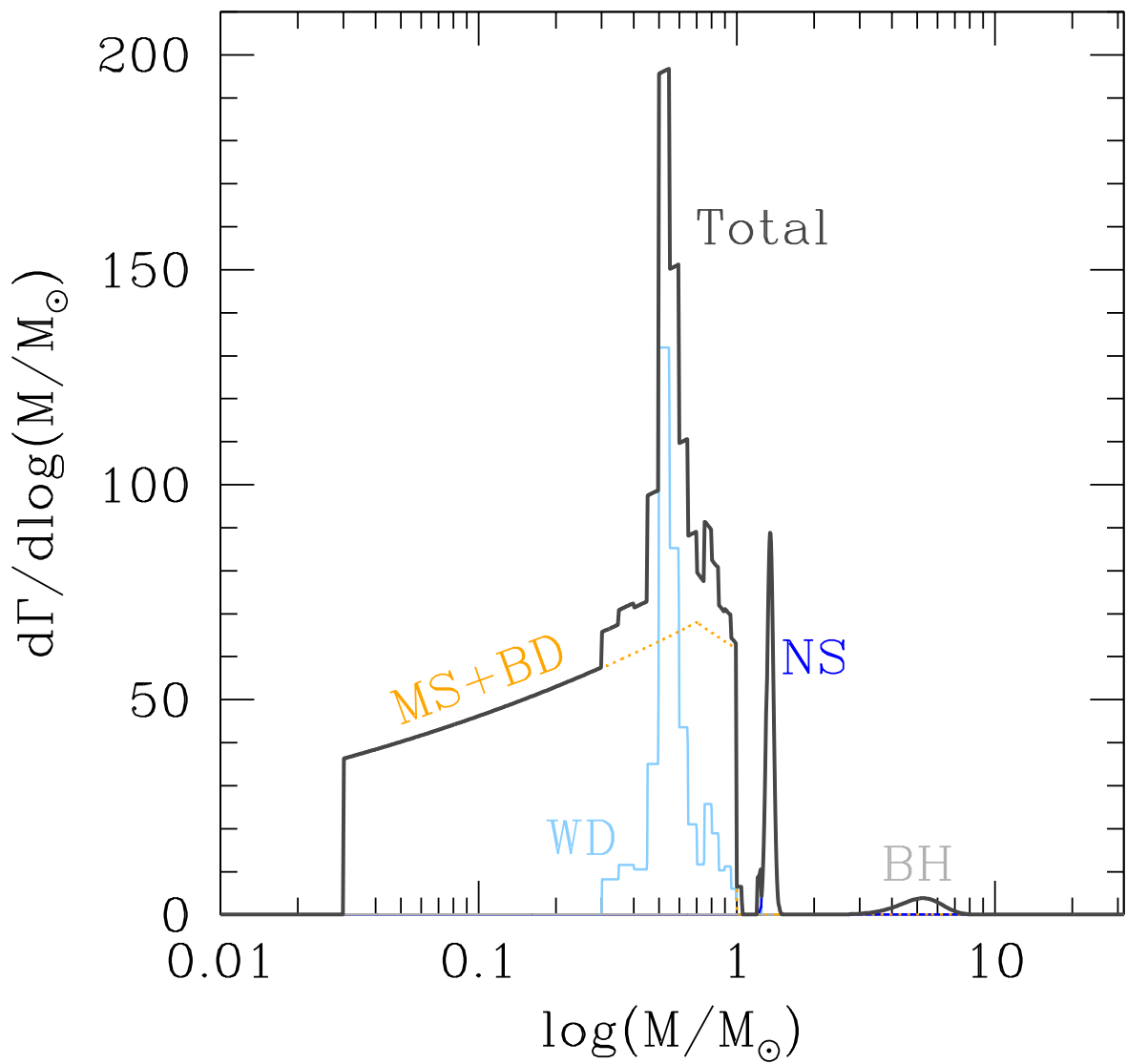}
\caption{The solid black curve shows
the theoretical estimate of total event rate of microlensing
events toward the Galactic bulge from \citet{gould2000}, as a function of the mass
of the lens.  This event rate is decomposed into its contribution
from main-sequence and brown dwarf lenses,
white dwarfs, neutron stars, and black holes, as indicated.  The sensitivity
of microlensing planet surveys 
is roughly proportional to this event rate. Adapted from \citet{gould2000},
reproduced by permission of the AAS.
\label{fig:mass}}
\end{figure}

\end{itemize}

\subsection{Habitable Planets}

Of particular interest is the detection of terrestrial planets in the
Habitable Zones of their parent stars.  In \S\ref{sec:order}, we
discussed the magnitude of the signal expected for an Earthlike planet
separated by one AU from a solar type star for various detection
methods. These signals are generally quite small.  However, we can and
should also consider the signals and detectability of Earthlike
planets orbiting in the Habitable Zones of main-sequence stars with
mass significantly different than the sun.  Because the masses, radii,
and luminosities of stars on the main sequence can span an order of
magnitude or more, both the location of the Habitable Zone and the
magnitude of the signal expected from an Earthlike habitable planet
can vary substantially.  

In this subsection, we derive the scaling of
the detectability of Earthlike habitable planets with host star mass.
To this end, we adopt a power-law form for the stellar mass-bolometric
luminosity relation,
\beq
\frac{L_*}{L_\odot} \sim \left(\frac{M_*}{\Msol}\right)^4.
\label{eqn:ml}
\eeq
Our adopted exponent of 4 roughly corresponds to that found from a
weighted fit to the data for benchmark binary systems
analyzed in \citet{torres2010}.  This form provides a reasonable approximation of this
data set over the full span of masses of $\sim 0.2\Msol - 50 \Msol$.
We note, however, that there are significant deviations from this form within the range, particularly
for stars below the fully-convective boundary, which are generally
more luminous that predicted by Equation \ref{eqn:ml}.  Nevertheless, we will adopt
this simple form for the purposes of illustration.  With this adopted mass-luminosity
relation, we can define the location of the Habitable Zone as,
\beq
a_{\rm HZ}={\rm AU}\left(\frac{L_*}{L_\odot}\right)^{1/2} \sim {\rm AU}\left(\frac{M_*}{\Msol}\right)^2, 
\qquad P_{\rm HZ} \sim {\rm yr}\left(\frac{M_*}{\Msol}\right)^{5/2},
\label{eqn:ahz}
\eeq
Therefore, the Habitable Zone location ranges from a semimajor axis of $\sim 0.01$AU
and period of $\sim 1$ day for a star at the bottom of the main sequence to $\sim 4$ AU 
and $\sim 6$ years for a $M_*\sim 2~M_\odot$
star.  We adopt a mass-radius relation of the form,
\beq
\frac{R_*}{R_\odot} \sim \frac{M_*}{\Msol}
\label{eqn:rm}
\eeq
which describes the data in \citet{torres2010} reasonably well
for non-evolved stars with $M_* \la 2\Msol$.   

With these assumptions, we can now use the results from \S\ref{sec:sens} to
derive the scaling of the signal of a habitable planet with host star
mass for the various methods we have discussed.  

\begin{itemize}

\item {\bf Radial Velocities}. For radial velocity surveys, the radial velocity
and signal-to-noise ratio
for planets in the Habitable Zone scales as 
\beq
(\snr)_{RV} \propto M_p M_*^{-3/2}\qquad {\rm (Habitable)}.
\label{eqn:snrrvhab}
\eeq
Therefore, all else being equal, Habitable Zone planets
are significantly easier to detect around lower-mass stars. In particular,
for stars with $M\la 0.2~M_\odot$, the radial velocity amplitude for 
an Earth-mass planet in the Habitable Zone is expected
to be $\ga 1$m/s, which is within the reach of current
instrumentation \citep{bean2010}. 

\item {\bf Astrometry}. The astrometric signal and signal-to-noise ratio
for habitable planets scales as,
\beq
(\snr)_{AST} \propto M_p M_* d^{-1}\qquad ({\rm Habitable}),
\label{eqn:snrasthab}
\eeq
and thus at fixed distance and planet mass, astrometric surveys are more sensitive to habitable
planets orbiting higher-mass stars, provided that the period of the planets
is less than the duration of the survey.  
In addition, higher-mass stars are more luminous and thus have 
smaller photon noise uncertainties.  On the other hand, 
higher-mass stars are also less common and thus are typically more
distant.  The net result of these factors is that A and F stars
are the most promising targets for astrometric searches for planets
in the Habitable Zones of nearby stars \citep{gould2003}.

\item {\bf Imaging}.  For direct detection of habitable planets
in thermal equilibrium with their host stars, the planet star flux ratio and signal-to-noise ratio
scale as,
\beq
(\snr)_{dir} \propto R_p^2 M_*^{-4} \qquad ({\rm Reflected, Habitable}),
\eeq 
\beq
(\snr)_{dir} \propto R_p^2 M_*^{-5/2} \qquad ({\rm Thermal, Habitable}),
\eeq
strongly favoring low-mass stars.   Note that, by definition, the amount of stellar irradiation for a planet in the habitable zone is independent of the mass of host star, and thus for fixed planet properties the thermal or reflected flux of the planet is also independent of the mass of the host star.  The dependence on stellar mass in the above scaling relations therefore arises simply from the change in the flux of the star. However,
the second requirement for direct detection is that the angular separation
of the planet from its parent star must be larger than the inner working
angle of system.  At fixed mass, this translates into a maximum distance
that a Habitable Zone planet can be detected,
\beq
d_{max} = 10~{\rm pc}\left(\frac{\theta_{{\rm IWA}}}{100~{\rm mas}}\right)^{-1}
\left(\frac{M_*}{\Msol}\right)^2\qquad ({\rm Habitable}).
\label{eqn:dmaxhab}
\eeq
The number of available targets is $\propto n(M_*)d_{max}^3$, where
$n(M_*)$ is the volume density of stars of a given mass, i.e., the mass
function. Since the exponent of the mass function in the local
solar neighborhood is generally $\ga -2$, this requirement strongly favors 
high-mass stars.  The optimal mass will depend on the precise details
of the survey and the nature of the noise sources (see, e.g., \citealt{agol2007}),
but these arguments demonstrate that we can generically expect the
sensitivity of imaging surveys for habitable planets
to be fairly strongly peaked at intermediate masses.

\item {\bf Transits}. The signal-to-noise ratio, transit probability,
and period of a transiting habitable planet scale as,
\beq
(\snr)_{tr} \propto R_p^2 M_*^{-5/2}, \qquad P_{tr} \propto M_*^{-1}, \qquad P \propto M_*^{5/2},
\qquad ({\rm Habitable})
\label{eqn:trhabitable}
\eeq
all of which favor or strongly favor low-mass stars.  Furthermore, as discussed above,
the radial velocity signals of Habitable Zone planets around low-mass stars are also
larger and within reach.  Finally, low-mass stars are more common.  
These various considerations have led to the suggestion that transit surveys of
low-mass stars may be the most promising route to detecting habitable Earthlike
planets \citep{gould2003b,nutzman2008,blake2008}.  Indeed, several such surveys are underway
or are being planned (e.g., \citealt{charbonneau2009}), with the ultimate goal
of finding a Earthlike system whose atmosphere can be characterized with, e.g.\ the
James Webb Space Telescope \citep{deming2009}.

\item {\bf Microlensing}.  The system parameters which determine the detectability
of a given planetary system with gravitational microlensing are the mass
ratio $q$ and projected separation $s$ in units of $r_\e$.  For a habitable Earthlike
planet these are,
\beq
q \sim 5\times 10^{-5} \left(\frac{M_p}{M_\oplus}\right)\left(\frac{M_*}{0.5 \Msol}\right),
\label{eqn:qearth}
\eeq
\beq
s_{\rm HZ} \equiv \frac{a_{HZ,\perp}}{r_\e}
\sim 0.1 \left(\frac{M}{0.5 M_\odot}\right)^{3/2}
\left(\frac{d_s}{8~{\rm kpc}}\right)^{-1/2}
\left[\frac{x(1-x)}{0.25}\right]^{-1/2}, \qquad ({\rm Habitable})
\label{eqn:shz}
\eeq
where in the expression for $s_{\rm HZ}$ we have assumed a median projection factor
such that $a_{HZ,\perp} = 0.866 a_{\rm HZ}$.  Therefore, for typical microlensing host
stars, the Habitable Zone distance is much smaller than the Einstein ring. 
While mass ratios of $q\sim 10^{-5}$ are
readily detectable for planets with separations near the Einstein ring ($s \sim 1$), they 
are much more difficult to detect for planets with $s\ll 1$. This is because these 
such planets can only be detected when they perturb the inner image created by 
the primary host star, and then only when this image is close to the primary and thus 
highly demagnified. See Figure \ref{fig:cartoon}.  These perturbations are therefore generally quite
small.  Furthermore, perturbations of the inner image are more strongly suppressed
for large source stars \citep{gould1997,bennett1996}.  

From Equation \ref{eqn:shz}, 
we see that microlensing favors the detection of habitable planets around higher-mass
stars \citep{distefano2008}, and stars that are close to the source or close to the Earth (i.e., such
that $x(1-x)$ is small).  While current and next-generation ground-based microlensing surveys
have essentially no sensitivity to habitable Earthlike planets, specialized surveys for nearby
microlensing events, or space-based surveys which boast much larger event rates and 
detection efficiencies, could potentially detect such systems 
\citep{distefano2008,park2006,bennett2010,green2011}.

\end{itemize}

\section{Early Milestones in the Detection of Exoplanets}
\subsection{Van de Kamp and Barnard's Star}
The pre-1995 literature is scattered with several (presumably) spurious
claims of detections of planets around nearby stars.  Perhaps the best
known early claim is that of van de Kamp, who conducted a astrometric
campaign to detect ``dark'' companions to nearby stars \citep{vandeKamp}.  Van de Kamp's
lower limits were impressive, typically ruling out Jupiter-mass objects
in periods of years to a couple decades, and he reported several stars
as having barely-detectable companions of apparently substellar mass.
Most intriguing was his report of first one, then later two
Jupiter-mass companions to Barnard's star (GJ 699), the second closest stellar
system to Earth.

Van de Kamp made astrometric measurements from the positions of the
apparent centroids of stellar images on photographic plates, and targets such as
Barnard's star suffered from having a constantly changing set of
astrometric references over his multi-decade survey due to its
record-high proper motion (over 10''/yr).   Subsequent astrometric and
radial velocity work have ruled out his claims (Choi et al. 2012, ApJ, submitted, and
references therein).

\subsection{PSR 1257+12 and the Pulsar Planets}

Pulsars are exquisite clocks, typically producing pulses with periods
of order $\sim 1$--$10^{-3}$ s.  Once these periods are corrected for
well-measured linear drifts with time and occasional ``glitches''
(sudden shifts in period), their precision can rival and even surpasses the best atomic
clocks on Earth.   Successful analysis of pulse arrival times requires
carefully solving for the distance and space motion of the pulsar and
the removal of the effects of the motion of the observatory.  

In 1991, two teams announced having contemporaneously observed
unexplained residuals to their timing models indicative of the first
known exoplanets: very small, terrestrial-planet-mass companions
orbiting their pulsars.  Matthew Bailes and Andrew Lyne \citep{Bailes1} reported a timing
variation with a period of 6 months apparently due to a 10 \mearth\
companion orbiting the pulsar PSR 1829-10.  Meanwhile, Cornell
astronomer Alexander Wolszczan had observed a 
similar sort of signal from a millisecond pulsar, PSR 1257+12, and had recruited
Dale Frail of the National Radio Astronomical Observatory to help
confirm its position on the sky to perfect the position model.   By
November 1991 \citet{Wolszczan92} had submitted a manuscript
on their discovery, and both teams planned to
describe their work at the January 1992 meeting of the American
Astronomical Society.

At the meeting, Lyne announced that just
days earlier he had discovered an error in his timing model.  A
tiny positional error combined with an insufficiently precise description
of the Earth's orbit had led to the small, periodic, 6-month signal in their residuals that they
had mistaken for a planet.  With the correct timing model, there was
no evidence of a planetary perturber on the pulsar.  Lyne's
public and frank admission was acknowledged as a laudable
demonstration of scientific integrity by a standing ovation at the
meeting and an editorial in {\it Nature} \citep{Bailes2,Nature92,Wolszczan12}.  

The world would not be long with its first exopanets,
however.  The very next speaker at the AAS meeting was Wolszczan,
whose timing model had correctly accounted for all important effects.
Wolszczan described the first planets known outside the solar system:
a pair of bodies with $\sim 4$  \mearth\ orbiting the
millisecond pulsar PSR 1257+12 with periods of 66 and 98 days.  This
system would continue to impress, revealing a third low mass planet to
Wolszczan's team, as well \citep{Wolszczan00}.   

These planets' formation mechanism is still not understood, and to
date no similar system of low mass planets is known.  Signals of
higher-mass planets orbiting pulsars would continue to be found,
however: in particular, Bailes would go on to discover an apparently
high-density 1.4 \mjup\ object orbiting pulsar PSR J1719-1438 \citep{Bailes11}.

\subsection{Early Radial Velocity Work}
\subsubsection{Campbell \& Walker's survey and $\gamma$ Cep A{\it b}}
\label{Campbell}
In the 1981, Bruce Campbell of the University of Victoria, Gordon
Walker of the University of British Columbia, and their team began an
ambitious survey for substellar 
objects orbiting 20 of the brightest nearby stars.  They sought to
exploit the recently available technologies of high resolution (R $>$
40,000) echelle spectrographs and digital detectors in the form of
Reticons and CCDs to achieve the best possible Doppler precision. 

To establish an
accurate and robust wavelength solution they employed an absorption
cell of hydrogen fluoride (HF) which imposed a ``picket  fence'' of
regular, strong, narrow, and widely spaced ($\sim 15$\AA) absorption lines from the $3
\rightarrow 0 R$ branch transitions near 8700\AA, far from any
telluric features \citep{Campbell88}.   Campbell \& Walker had sought
a chemical cell that would provide such lines in the optical that
would not contaminate the stellar lines, and with few extraneous absorption
features from isotopic impurities or other effects.  Cambell \& Walker
found no other chemicals that fit their needs, and so used HF despite
the serious difficulties of working with that
chemical\footnote{Campbell \& Walker understatedly reported the ``obnoxious'' 
nature of HF and that ``standard safety precautions of chemical
laboratories are appropriate'' for this rather dangerous chemical which
is reacts with glass, forms hydrofluoric acid on contact with water,
can painlessly penetrate human tissues, and causes burns that can
necessitate amputation (OSHA Occupational Safety and Health Guidelines
for HF, US Department of Labor, 2012).}.  To achieve the necessary optical
depth and wavelength stability of HF lines, Campbell \& Walker
stabilized the temperature and pressure of their 1 m long cell at 373 K and connected it to
a vessel containing liquid HF in an ice bath, yielding a pressure of roughly 0.5 atm.

The well-known and well-spaced HF lines granted Cambpell \& Walker
unprecedented optical Doppler precision, independent of mechanical and
thermal changes in the optics of their spectrograph.  This method was,
at the time, greatly superior to emission line calibration because the
reference lines were measured simultaneously to the stellar exposure
and through the same optical path as the stellar photons.  Cambell \&
Walker achieved 10--15 m/s radial velocity precision on most of their
sample of 20 stars.  This precision is, in retrospect, more than
sufficient to detect close-in giant exoplanets, but their sample of stars was simply
too small and close-in planets are too rare for them to have discovered a
strong exoplanetary signal with any significant  likelihood.

\citet{Campbell88} were able to confirm that objects with $\msini=10$--$80$ $\mjup$
are rare, and noted several interesting signals near the limits of
their detectability.  They noted an especially intriguing apparent signal
for $\gamma$ Cep A with a 25 m/s amplitude and 2.7 yr period,
superimposed on a long-term acceleration from the star's binary
companion.  The implication of this periodicity was that $\gamma$ Cep
A was orbited by a $\sim$ 1.7 \mjup\ mass planet at a few AU. 

Four years later, \citet{Walker92} reported on their monitoring of
the Ca {\sc ii} 8662\AA\ line and determined that $\gamma$ Cep
had a weak 2.52 yr activity period, uncomfortably close to the purported
planetary signal, and cautiously noted that the RV signal was likely
due to stellar activity.  

Eleven years later, after nearly 100 bona fide exoplanets had been
discovered by teams around the world, \citet{Hatzes03} announced
RV monitoring at McDonald Observatory
had confirmed Campbell, Walker, \& Young's original detection:  the
variations were indeed due to a $\msini = 1.7$ \mjup, $P=2.5$ yr
planet, and there was no longer any evidence of a 2.5 yr activity
period.  In retrospect, Campbell \& Walker's planet search had been a
sort of success, after all.  It had also inspired subsequent teams to
continue their efforts.

\begin{figure}
\plotone{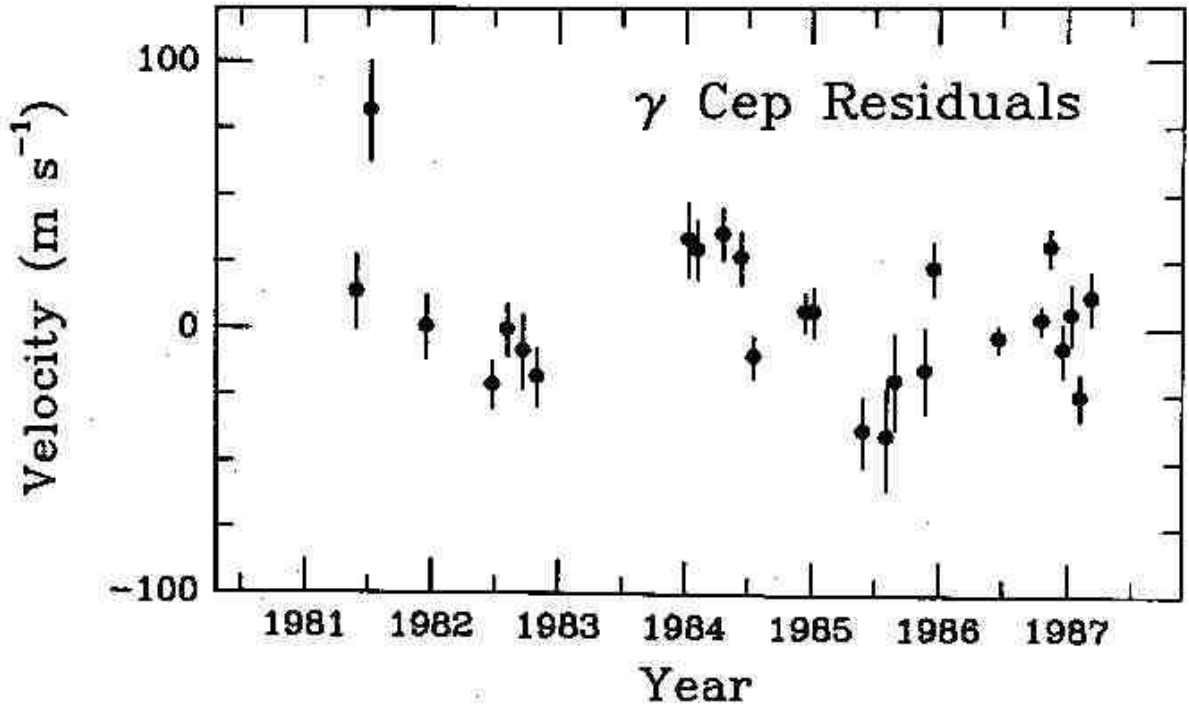}
\caption{Figure from \citet{Campbell88} illustrating the first
  tentative detection of a real exoplanet from the pioneering radial
  velocity survey. Reproduced by permission of the AAS.}
\end{figure}

\begin{figure}
\plotone{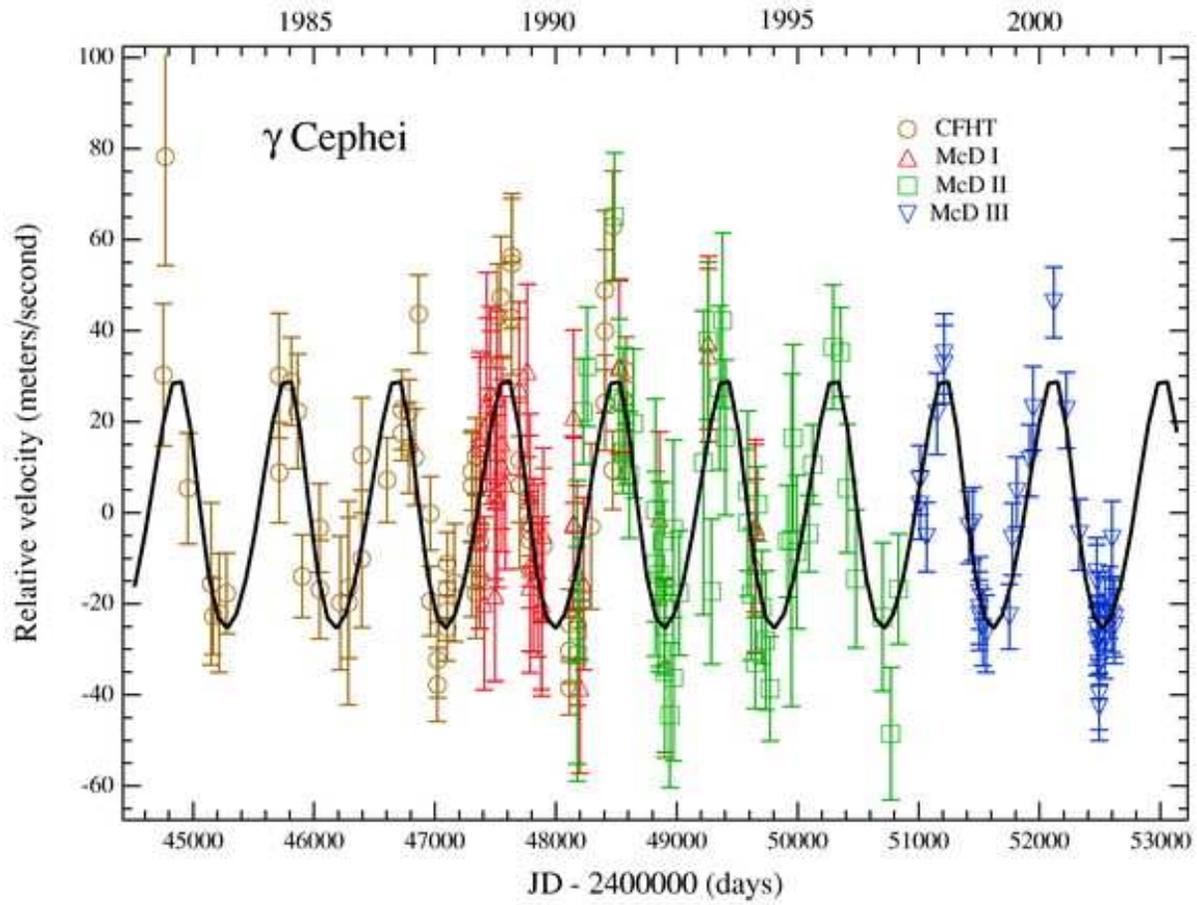}
\caption{Fig.~3 of \citet{Hatzes03} showing confirmation data for the
  planet $\gamma$ Cep $b$, including the original
  \citeauthor{Campbell88} data.  Reproduced by permission of the AAS.}
\end{figure}

\subsubsection{Latham's survey and HD 114762 {\it b}}
In 1988, David Latham of the Harvard-Smithsonian Center for Astrophysics and his team
described a new result from their precise radial velocity
instrument, the CfA Digital Speedometer at Oak Ridge Observatory.  In 1984, Latham and Israeli astronomer Tsevi Mazeh had conducted a short survey of about three dozen early M dwarfs for short period, massive planets using this instrument, but had found nothing \citep{Latham12}.   For this new survey, Latham et al.\ sought to
further stabilize their spectrograph to achieve $\sim$ 100 m/s
precision to measure accurate binary orbits and improve the IAU system
of radial velocity standards.  Latham et al.\ achieved this stabilization by
removing the Cassegrian instrument from the telescope, stabilizing its
temperature, and feeding it with a 100-$\mu$ optical fiber.  This
provided a constant gravity vector, thermal and mechanical stability,
and a (relatively) uniformly illuminated entrance slit, robust against
guiding errors.  

On its first night of operation, \citet{Latham89} observed several RV standard
stars including HD 114762.  They noted a large (390 m/s) radial
velocity discrepancy from the known value.  Curious, they compared
their previous measurements made at lower precision and found a highly
significant signal at near 84 d with 530 m/s semiamplitude,
corresponding to a 13 \mjup\ companion.  

\begin{figure}
\plotone{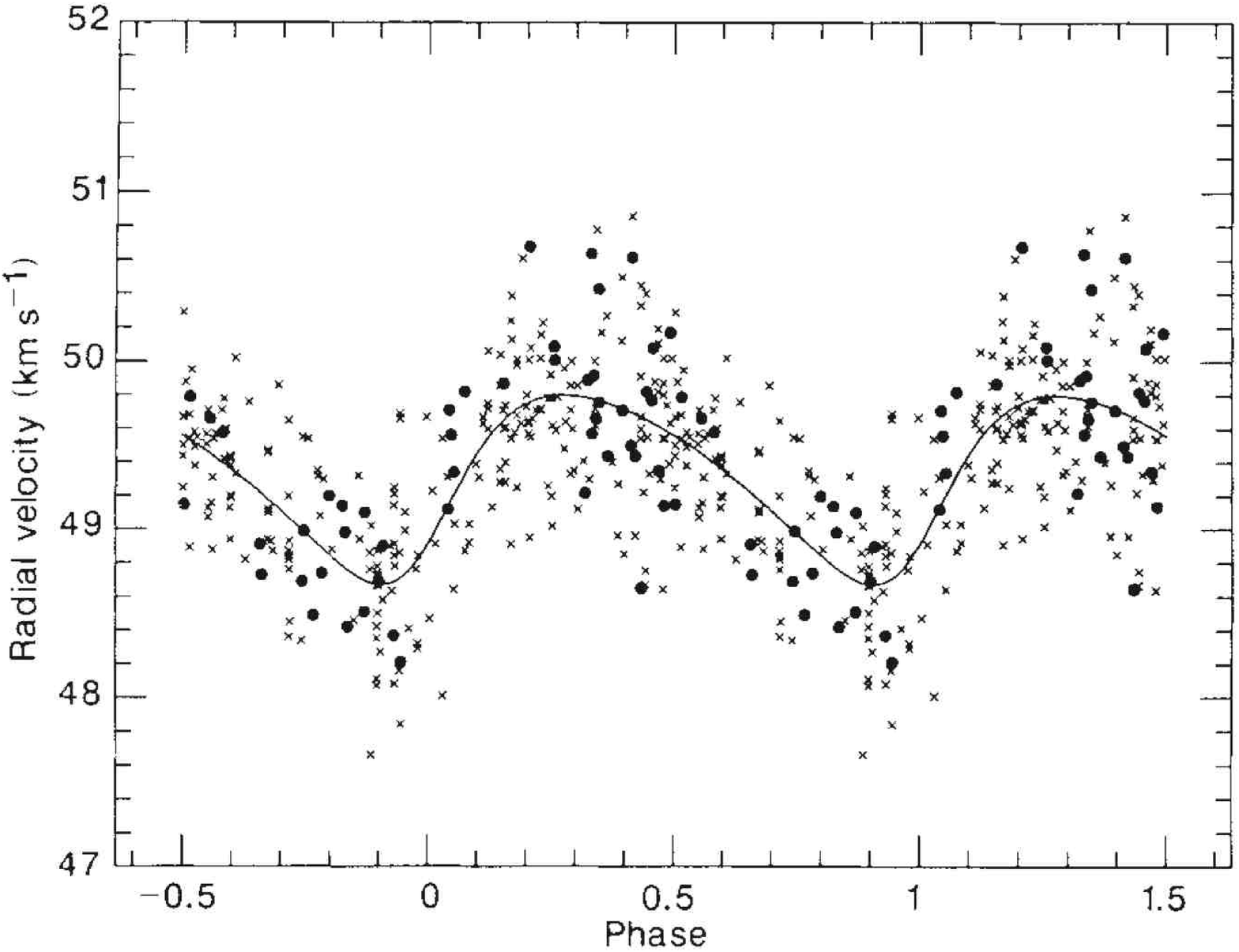}
\caption{Figure 2 of \citet{Latham89} illustrating the first firm
   detection of a substellar object outside the Solar System.  In retrospect,
   the object appears to occupy the high-mass tail of the
   distribution exoplanets.  Figure includes both CORAVEL (filled
   circles) and CFA Digital Speedometer (crosses) measurements.\label{Latham}}
\end{figure}

Subsequent observations at high precision confirmed the reality of the
signal (see Fig.~\ref{Latham}).  \citet{Latham89} had discovered, serendipitously, on the
first night of observation, and in a sample of only seven objects,
what today would be considered by many to be an exoplanet.  At the time, no objects
with a mass anything like 13 \mjup\ were known, and Latham et al.\ cautiously
referred to their object as ``a probable brown dwarf'', however they noted that the companion ``might even be a giant planet''\footnote{Later, the IAU would provisionally set the
  deuterium-burning limit for star-like objects, 13 \mjup\, as the border
  between planets and brown dwarfs found in orbit around stars.  This
  is useful for purposes of nomenclature,
  but bears little on issues of formation and composition,
  and today the distinction is not always honored.}, a point that was picked up by the media but criticized by many of their colleagues \citep{Latham12}.  Regardless of its taxonomic class, it was the first firm detection
of a substellar object beyond the Solar System, and today is often
included in catalogs of exoplanets \citep[e.g.][]{EOD}.

\subsubsection{Marcy \& Butler's iodine survey}

In 1992, Geoffrey Marcy and Paul Butler of San Francisco State University announced
their survey of 70 nearby stars 
using an iodine (I$_2$) cell for wavelength calibration.  Previous
Doppler work by \citet{Marcy89} had used ThAr emission lamps to search
for brown dwarfs orbiting nearby M dwarfs with radial velocity
precision of $\sim$ 250 m/s;  Marcy \& Butler sought to improve this
precision by 1--2 orders of magnitude with their iodine cell and the
CCD at the high-resolution Hamilton spectrograph, designed by \citet{Vogt}. 

Unlike Campbell \& Walker, Marcy \& Butler sought an absorption cell
that would provide absorption features {\it throughout} a broad region
of wavelength space.  Following \citet{Libbrecht88} (who had employed an iodine
cell to study solar sunspots and had extended their studies to
p-mode measurements of stars) Marcy \& Butler settled on iodine as the ideal
absorption gas.   Their rationale was that this provided
the wavelength reference at every point in the spectrum, not just every
15\AA, and that the potentially problematic blending of stellar
features with iodine features could be modeled given a sufficiently
accurate template spectrum of the star and the iodine cell (the latter
obtained with a Fourier 
Transform Spectrograph).  

Marcy \& Butler further sought to model the variable instrumental
profile of the spectrograph as a function 
of wavelength to account for the not insubstantial thermal and
mechanical variations in the spectrograph, and for the non-uniform
illumination of the entrance 
slit.   Marcy \& Butler's sealed cell was $\sim$ 10 cm long, held at 
a constant 323 K, and 0.01 atm, and had the further advantage that it
was relatively easy to construct and its contents were benign (indeed,
medicinal!).
\citet{Marcy92} reported that they had 
achieved 25 m/s precision at the beginning of their survey, and
foresaw significant improvement through more sophisticated
instrumental profile modeling.  Indeed, by 1996 Marcy \& Butler would
demonstrate 3 m/s 
precision \citep{Butler96b} and they and their collaborators would go
on to be responsible for over half of the exoplanets discovered over
the next 15 years. 

\subsubsection{Hatzes \& Cochran's survey and $\beta$ Gem {\it b}}

\citet{Hatzes93} reported their results from precise Doppler
monitoring of three bright K giants as part of a broader planet
detection effort.  Their primary technique was to use telluric
(atmospheric) O$_2$ bands as an absorption wavelength reference, which
had been reported by \citet{Griffin73} to be sufficiently stable to
  allow 10 m/s precision.  They found that typical long-term
  stabilities were more like 20 m/s.   They had also begun
  employing an iodine cell \citep{Cochran93}, and at this point had
  obtained a small number of iodine observations.

Hatzes \& Cochran found that all three K giants in their sample,
Arcturus ($\alpha$ Boo), Aldebaran ($\alpha$ Tau), and Pollux ($\beta$
Gem) displayed large, periodic radial velocity variations with
semiamplitudes of 50-200 m/s.  Comparison with prior radial velocities
obtained by Cambpell's group \citep{Walker89} revealed that the
variations were coherent over 10 years.  While both $\alpha$ Boo and
$\alpha$ Tau showed significant day-to-day RV variations indicative of
radial pulsation modes and correlated variations in the
10830\AA\ He {\sc i} line, $\beta$ Gem seemed to have a clean signal,
consistent with a 554 d planet with a minimum mass of 3 \mjup. 

Observations by the Canadian team \citep{Larson93} showed that the
Ca {\sc ii} 8662\AA\ line showed periodic variation at the same
frequency as the RV variations (which they had measured
independently).  This coincidence cast strong doubt on the planetary 
interpretation of the $\beta$ Gem RV variations, especially in light
of the much larger and more clearly activity-related variations in
$\alpha$ Boo and $\alpha$ Tau.  

\citet{Hatzes06} combined literature data with subsequent iodine
observations from McDonald Observatory and Tautenburg Observatory to
show that the RV variations continued coherently into 2006 and that
the Ca {\sc ii} H \& K lines showed no coincident variation.  They
concluded that the variations in $\beta$ Gem were likely due to a
minimum mass 3 \mjup\ companion with period 590 d.  

\subsubsection{Mayor \& Queloz and 51 Pegasi {\it b}}
\label{Mayor}
The first unambiguous detection of a planet-mass object orbiting a
normal star was by \citet{Mayor_queloz} of Geneva Observatory.  Michel Mayor 
and Didier Queloz used the
ELODIE spectrograph, which achieved 13 m/s precision through outstanding
mechanical stability.  ELODIE (the successor to CORAVEL) was a fiber
fed spectrograph within a stable, temperature controlled environment
\citep{Queloz98}.    Wavelength calibration was achieved through use of
a simultaneous observation of a thorium-argon (ThAr) emission lamp.
Mayor \& Queloz used cross-correlation with a binary mask to determine the velocity of
the stellar spectrum with respect to the known wavelengths in the
emission line spectrum.  The mechanical stability of 
the instrument ensured that the offset between the stellar and
comparison lamp spectra was fixed and stable, and the scrambling
inherent to the fiber ensured that the position of the stellar
spectrum did not suffer significantly from variations in illumination
or guiding on the fiber tip.   

51 Peg $b$ has a semiamplitude of only 59 m/s and period of only 4.2 d,
implying a minimum mass of 0.5 \mjup.  This was a shocking development
--- planet formation theory had not predicted the existence of such
close-in planets\footnote{But see the remarkably prescient article by
  \citet{Struve52}, which all but foresaw this detection, how it would
  be made, and the subsequent detection of planetary transits.}, and indeed the tiny mass 
implied by the detection was smaller than 
any known binary companion by more than an order of magnitude.\footnote{The
  third firm detection of a substellar object, the imaging of brown dwarf GJ 299 B, was announced at
  the same conference as 51 Peg $b$!}  Immediately \citet{Marcy95}
confirmed the detection, as did \citet{Hatzes97} soon thereafter.
Debate ensued about the nature of the variations and whether they
could be due to non-radial pulsation modes \citep{Gray97}, but the
detection of planetary transits would put these concerns to
rest:  the field of exoplanetary science had begun in earnest.  The
Geneva team would expand and find great success over the next decades,
eventually pushing their precision below 1 m/s with the HARPS 
spectrograph.

\subsection{The First Planetary Transit: HD 209458{\it b}}

The presence of close-in planets
provided an opportunity to detect exoplanets directly through
transits.  The probability that a planet will transit its hosts star
is inversely proportional to its orbital distance, and since 51 Peg $b$
and similar ``Hot Jupiters'' orbited at $\sim 10$ stellar radii from
their host stars, their transit probability was around 10\%.
Photometrists began to monitor these planets' hosts stars for such
events, expecting to find one once the number of known systems
approached 10.  Concerns over nonradial pulsations also contributed to
the desire to monitor stars for photometric evidence of such effects.

Two teams detected the $\msini = 0.7$ $\mjup$, $P=3.5$ d planet orbiting
HD 209458 independently \citep[]{Mazeh00,HenryG00} and collaborated with
photometrists to conduct the now-standard photometric follow-up prior
to publication.  Two teams succeeded contemporaneously: their
announcements of the detection of the transits of HD 209458 appeared in the 
literature simultaneously, having been submitted to the Astrophysical Journal
within one day of each other \citep{HenryG00,Charbonneau00} exemplifying
the intense competition to produce exoplanetary ``firsts''.  This
measurement of the orbital inclination and radius (and thus the true
mass and density) of the planet dispelled any remaining doubt as to the origins of
most of the similar RV variations of stars, and provided the necessary impetus for large-scale
efforts to detect more planets with radial velocities, transits, and,
soon, microlensing and direct imaging.

\subsection{Microlensing}

\subsubsection{Microlensing History}

While the idea of gravitational microlensing by individual stars was
considered sporadically over the past century  
\citep{einstein1936,eddington1920,chwolson1924,lodge1919,liebes1964,refsdal1964},
it was the seminal paper by \citet{paczynski1986} that gave birth to
the modern microlensing field.  In this paper, Paczy{\' n}ski argued
that it would be feasible to monitor several million stars toward the
Magellanic clouds on timescales of a few hours to a few years, in
order to search for gravitational microlensing events due to
foreground massive compact objects that could make up a substantial
fraction of the mass of the dark matter halo of the Milky Way.  Within
a few years, several collaborations were initiated to survey regions
in the Large Magellenic Clouds and Galactic bulge to search for
microlensing events \citep{alcock1993,aubourg1993,udalski1993}.  The
first detections followed shortly thereafter, and to date of order
$10^4$ microlensing events have been detected, with the majority seen
along the line of the sight to the bulge.

Although the original motivation for microlensing surveys was the
search for dark matter, it was soon realized that it would be possible
to search for planetary companions to the stars and remnants that
provided a guaranteed signal for these experiments.  \citet{mao1991}
first pointed out that binary lenses whose components were separated
by roughly their Einstein ring radius would give to sharp, distinctive
light curves features associated with the presence of caustic curves
in such systems.  Caustics are the set of source positions where extra
image pairs are created or destroyed with the source crosses the
caustic, resulting in large changes in the total magnification.  They
also noted that the probability of a source crossing these caustics
for a binary lens remained substantial down to mass ratios of $q\sim
10^{-3}$, therefore suggesting that planetary companions could also be
detected in this way.  \citet{gl1992} consider this idea in detail,
refining the estimates for the detection probabilities for different
planet mass ratios, and discussing the practical requirements for
carrying out an exoplanet survey with microlensing.  In particular,
they advocated a two-tier strategy, whereby survey collaborations use
a single dedicated telescope equipped with a wide-field camera monitor
large areas of the sky to identify and alert stellar microlensing
events before peak, and follow-up collaborations with access to
several longitudinally-distributed, narrow-angle telescopes follow
particularly promising events at much higher cadence to search for
the brief planetary deviations.  The first microlensing planet surveys began in 
1995, with the first real-time alerts from the survey
collaborations (e.g., \citealt{udalski1994,alcock1996}), and subsequent
monitoring of these alerts by several follow-up collaborations 
\citep{alcock1996,albrow1998,rhie2000}.  

Ongoing surveys over the next 6 years (1995-2001) failed to detect any
planetary microlensing events.  The primary reason for this is that
the total number of events alerted by the survey collaborations was
relatively low ($\la 100$), and therefore there were typically few 
events ongoing at any given time that were both suitable for
follow-up and very sensitive to planets.  In particular, there were
only a handful of high-magnification events per year, which had been
recognized to be intrinsically very sensitive to planetary companions
\citep{griest1998}.  Nevertheless, this phase was important for the
field, as the real-world struggles involved with carrying out and
analyzing the results from these surveys, including the ensemble of
non-detections \citep{gaudi2002,snodgrass2004}, naturally led to the
development and maturation of both the theory and practice of the
method.

\subsubsection{First Planet Detections with Microlensing}

The first detections of planets with microlensing were enabled
primarily by a series of upgrades by several survey collaborations to
their observational setups. In 2001, the OGLE collaboration initiated
their third phase with an upgrade to a new camera with a 16 times
larger field of view.  With this larger field of view, they were able
to monitor a larger area of the Galactic bulge with higher cadence,
and as a result began alerting $\sim 500$ microlensing events per
year.  This higher event rate, combined with improved cooperation
between the survey collaborations, led to the first planet discovery
in 2003 by
the Microlensing Observations in Astrophysics (MOA) and OGLE
collaborations \citep{bond2004}.  Shortly thereafter, the MOA collaboration upgraded to a 1.8m
telescope with a 2 deg$^2$ camera \citep{sako2008}.  By 2007, the MOA
and OGLE collaborations were sending alerts for nearly 1000 microlensing events
per year, thus enabling a substantial increase in the rate of planet
detections.

\begin{figure}
\plottwo{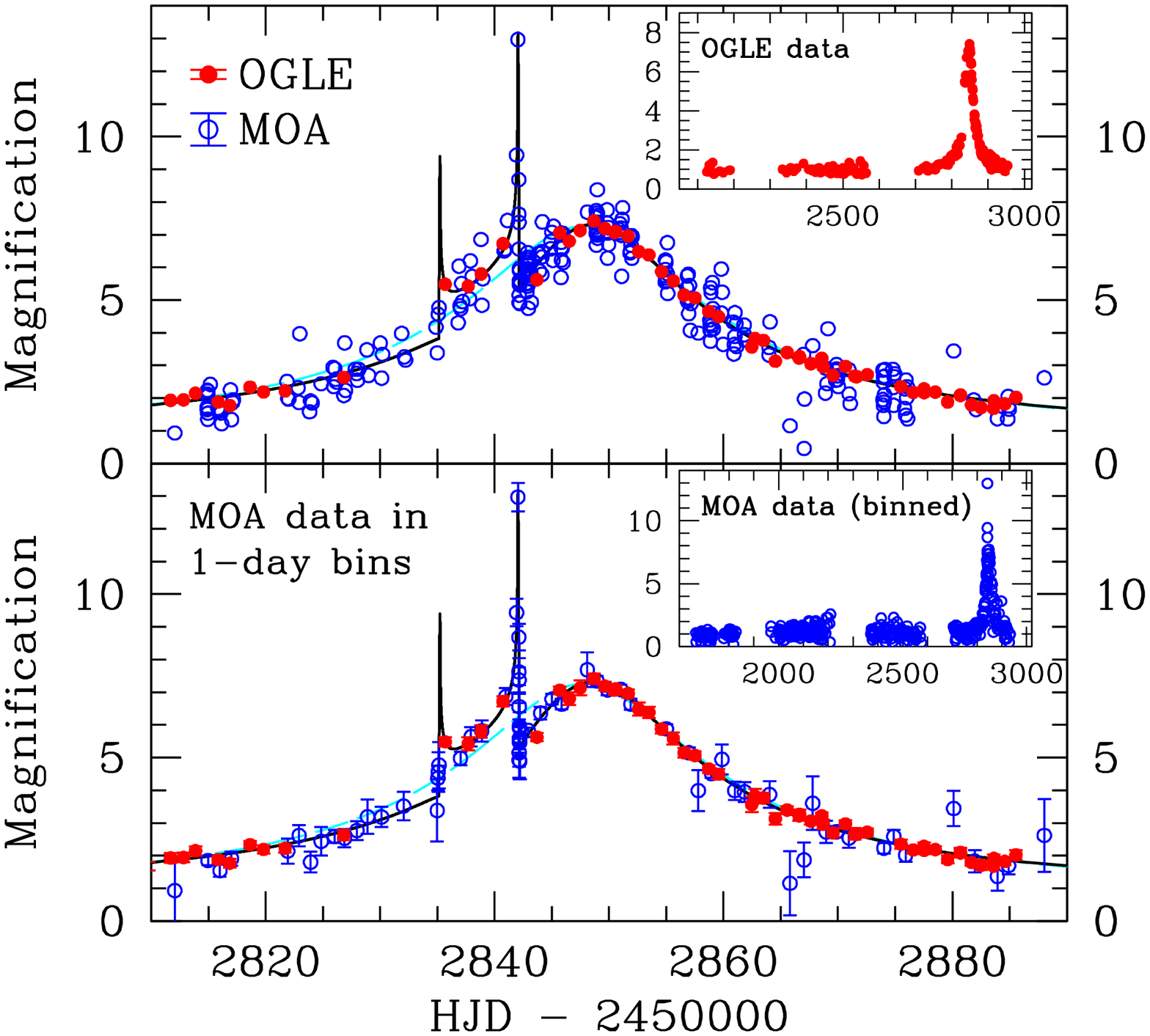}{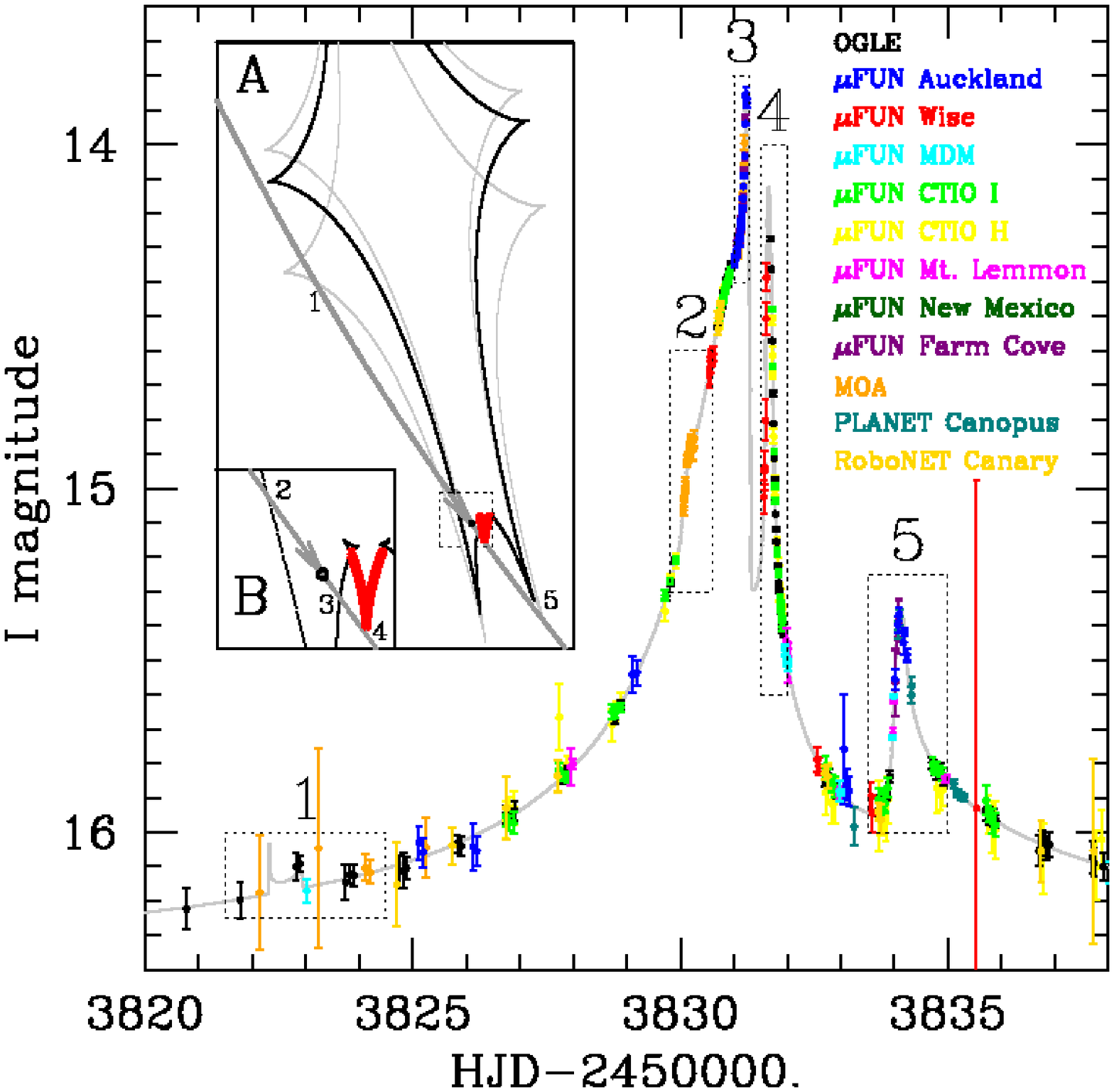}
\caption{
(Left) The first discovery of an exoplanet with microlensing 
in the OGLE 2003-BLG-235/MOA 2003-BLG-53 event \citep{bond2004}.
The red and blue points show the data from the OGLE and MOA collaborations,
respectively.  The top large panel shows the native data, whereas the MOA data
have been binned into 1 day bins in the bottom panel.  The black and cyan
curves show the best fit planetary and single lens model, respectively.
The planetary companion is revealed through the brief deviation
from the smooth symmetric curve arising from the host star, including the well-covered sharp spike
near HJD-2450000$\sim$2842 caused by the source crossing a caustic created by
the planetary companion.
The small insets show the full, multi-year data spans for the OGLE and MOA data.
From \citet{bond2004}, reproduced by permission of the AAS.
(Right) The OGLE-2005-BLG-109 microlensing
event, arising from a star with a Jupiter/Saturn analog two-planet system.  
Panel A shows the source trajectory through the caustic created by the two
planets (dark gray curve).  The five light curve features are caused by
the source crossing or approaching the caustic, and
the locations of these features indicated with numbers.  The majority of
the caustic (in black) is due the Saturn-analog
planet, which explains 4 of the 5 features.  The
portion of the caustic arising from the Jupiter-analog planet
is shown in red.  This additional caustic is required to
explain the fourth feature in the light curve.  The
light gray curves show the caustic at the time of features 1 and 5.
Panel B shows the detail of the source trajectory and caustic near the times of
the second, third, and fourth features.  From \citet{gaudi2008}.
\label{fig:mlcurves}}
\end{figure}

The light curve data and best-fit model for the first microlensing
planet discovery are shown in Figure \ref{fig:mlcurves}.  This is
a ``cold Jupiter'': the planet has a mass $M_p \sim 3~\mjup$ and
orbits a star
with $M_* \simeq 0.6~M_\odot$ at a separation of $a \sim 4$AU, or
$\sim 2.5$ times the snow line distance \citep{bennett2006}.  

To date, 14 microlensing planet detections have
been published.  The masses and
semimajor axes of these planets are shown in Figure
\ref{fig:exoplanets}, they span nearly three decades in mass from a
few times the mass of the Earth to several times the mass of Jupiter,
and are spread over a factor of $\sim 5$ in separation, centered at a
few times the snow line distance.  Notable among these detections are
the first discovery of a ``cold SuperEarth'' \citep{beaulieu2006}, and
the first discovery Jupiter/Saturn analog
\citep{gaudi2008,bennett2010a}.

\section{State of the Art}

\subsection{Astrometry}

Astrometric precision has improved considerably since van de Kamp's
work, and the first verifiable astrometric discovery of an exoplanet
appears imminent.  \citet{vB10} announced that their long-term
astrometric monitoring of the ultra-cool dwarf star vB 10 had revealed a $\sim 6
\mjup$ companion in a 9 month orbit, however subsequent followup with
radial velocities determined that the signal was spurious
\citep{Bean10,Anglada10,Lazorenko11}.  

More promisingly, \citet{Muterspaugh10} used an optical interferometer
to carefully measure the astrometric motions of binary stars, and
combined these measurements with radial velocities of the systems to
search for low mass companions to stars in tight binaries.  The
project concluded with six planetary candidates, including two ``high
confidence'' members that could prove to be the first astrometrically
detected exoplanets.  If real, these planets will put strong
constraints on planet formation theories in binary systems.

The astrometric detection of planets discovered by other means has
produced substantially more results, primarily because the approximate astrometric
signals are known from prior radial velocity work and so searches are
more efficient.  Most fruitful has
been work on nearby stars employing the {\it Hubble Space Telescope} Fine Guidance
Sensor, which is capable of precise astrometry of bright stars.
This has revealed some high-mass planet candidates from radial velocity
surveys to be binary stars in face-on orbits, and has revealed the
mutual inclinations of planets in multiplanet systems
\citep{Bean2007,Bean2009,McArthur2010}.

\subsection{Imaging}
\subsubsection{2M1207{\it b}}
\citet{Chauvin10} describe their survey of young, nearby stars for low
mass, possibly planetary companions using the ESO/VLT 8-m telescope 
equipped with the NACO AO system and infrared camera.  They began their survey in 2002 during
commissioning, and targeted, among other things, the lowest mass
members of known, nearby, young stellar associations.  This allowed
them to maximize the separation and contrast of companions and push
coronagraphy into the planetary-mass regime.

\begin{figure}
\plotone{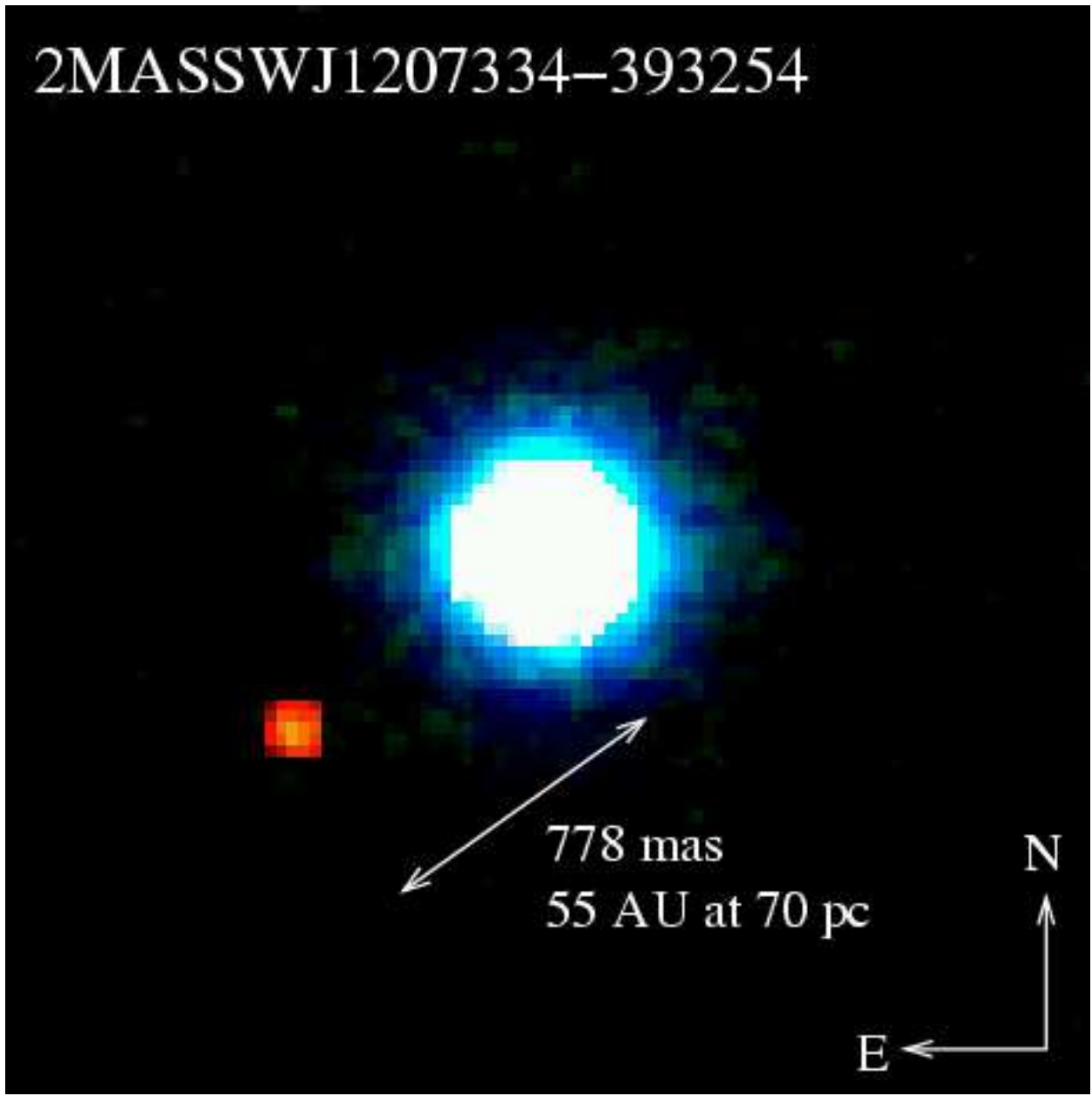}
\caption{NACO image of 2M1207$b$.  The primary is a M8 dwarf at $\sim
  70$ pc;  the secondary is much cooler late L dwarf with probable
  mass $\sim 5 \mjup$.  From \citet[Fig.\ 11][credit G. Chauvin \&
  ESO]{Chauvin04}. \label{1207}}
\end{figure}

After some promising detections of higher mass objects, the survey
bore fruit when \citet{Chauvin04} detected a very low mass companion
to the M8 TW Hydra association (TWA) brown dwarf 2MASSW J1207334-393254
(called 2M1207; Figure~\ref{1207}) at a separation of only 0\farcs8 ($\sim 55$ AU).  
Membership in the TWA yielded an age for the companion, and a distance was estimated
from the colors and brightness of 2M1207.  Comparison with models
of the thermal evolution of young objects \citep[][]{burrows1997,Chabrier00,Baraffe02} allowed \citet{Chauvin04}
to estimate the companions mass to be $\sim5~ \mjup$, assuming that it
was indeed a bound, coeval object and not a background contaminant.

\citet{Chauvin05} followed up their prior observations to confirm
common proper motion using the same instrument.  They demonstrated
that the two objects share proper motion and parallactic motion with
this star, all but proving that they form a bound pair. 

The nature of 2M1270$b$ was, and still is, unclear.  Its status as a
``planetary-mass'' object seems secure, but its wide separation (55 AU)
and $\sim$0.2  mass ratio with its primary made the
pair perhaps more analogous to a scaled-down binary star system than a
planet-star system.  \citet{Chauvin05} noted that a protoplanetary disk origin for
the $b$ component seemed unlikely.  Nonetheless, they had acquired the
first image of a planet, by mass if not by formation mechanism, and
this presaged the many more successes to come from high contrast imaging.
The ESO/VLT group would go on to detect the planet-mass object AB
Pic $b$ and many faint stellar companions to nearby stars, including
many planet hosts \citep{Chauvin10}.

\subsubsection{Fomalhaut {\it b}}
\citet{Kalas05} used the {\it Hubble Space Telescope} ({\it HST})  Advanced Camera for Surveys (ACS) to
image the dust belt orbiting Fomalhaut ($\alpha$ Piscis Austrini), a $\sim$400 Myr-old A4 dwarf at 7.7 pc.  
Debris disks or belts are a common feature of
many young main sequence stars, typically with structure consistent with
dynamical interactions from unseen planets (see chapter by Moro-Martin).  
The coronagraphic mode of ACS allowed the {\it HST} team to make a spectacular and detailed image
of the disk in reflected optical light, revealing that its dominant
feature is a highly inclined, off-center belt with an apparently sharp inner edge at 133 AU radius.   
The 15 AU geometric offset of the belt from the star and the sharp inner edge 
could be explained by the presence of a planetary companion with non-zero eccentricity
``sculpting'' the inner edge of the belt and maintaining the 
geometric offset via secular perturbation  \citep{wyatt99a}.

Follow-up observations to determine the structure of the disk using a
variety of PSF subtraction techniques allowed \citet{Kalas08} 
to confirm a persistent source with optical brightness
$\sim$ 25~mag and located $\sim$13$''$ away from the primary star, just inside the belt,
and consistent with the ``sculpting'' hypothesis.  Comparison of
multiple epochs allowed Kalas's team to measure the orbital motion of
the object astrometrically and confirm that it is a proper motion
companion of Fomalhaut.

Puzzlingly, this very faint companion did not appear in their
infrared imaging at the Keck 10-m and Gemini 8-m telescopes, which is
inconsistent with the optical emission being thermal in origin
according to current models.  \citeauthor{Kalas08} were able to
use the dust disk itself to constrain the object's mass dynamically to be $<3 \mjup$ \citep{Chiang09},
and thus the lowest mass directly imaged planet candidate to date. 
\citet{Kalas08} proposed several
possible explanations for Fomalhaut b's unusual optical brightness,
such as reflection from a circumplanetary dust ring that could be as
large as 35 planetary radii, though still significantly smaller than
Saturn's Phoebe ring.  They also proposed an alternative model where 
reflected light is due to a transient dust cloud produced by a rare destructive 
collision between two analogs of solar system Kuiper Belt objects.

As with 2M1207$b$, ascertaining the nature of Fomalhaut $b$ will
require additional observations and theoretical input.  It orbits an
intermediate mass star with semi-major axis $\sim 120$ AU, and is still
undetected in the infrared.  Whatever its nature, its presence in a disk of
material strongly implicates a disk origin for the object, and
provides hope for more secure similar detections of similarly bright
planetary objects in the future.

\subsubsection{Beta Pictoris $b$}

$\beta$ Pic is an A star with a prominent, edge-on debris disk
that was the first to be imaged in optical scattered light \citep{smith84}.  One unexpected 
result was that the disk had several asymmetries in
structure \citep{kalas95}, including a vertical warp in the disk midplane at $<$100 AU projected radius from the star \citep{burrows95}.  These and other
phenomenon observed toward $\beta$ Pic indirectly suggested the existence
of a planetary system, and $\beta$ Pic $b$ was finally directly imaged using
VLT/NACO \citep{lagrange09a}.  As with Fomalhaut, the debris disk structure could
be used to constrain the planet mass through dynamical theory \citep{mouillet97a}, as an
alternative to mass estimates based on planet luminosity models.  Its measured $L^\prime$ 
brightness of 11th mag corresponds to a $\sim8~\mjup$ planet at age $\sim$10 Myr.

$\beta$ Pic b is currently unique among the directly imaged exoplanets for having
the smallest semi-major axis, which, at $\sim$8 AU, corresponds to the approximate
ice-line of the system.  Unfortunately, the projected separation is also very small, 0\farcs4,
and follow-up spectroscopic study has yet to be obtained.  Systems such as 
$\beta$ Pic are therefore ideal targets for the next generation of extreme adaptive
optics instrumentation discussed below.  These future results will provide important
tests of various planet formation and luminosity evolution models, which in the 10 Myr-age
regime offer significantly different predictions for the physical properties of 
Jupiter-mass planets.

\subsubsection{The HR 8799 Planetary System}

A US/Canadian team led by Marois used AO coronagraphy on the Keck and
Gemini telescopes in angular differential imaging mode \citep[ADI,][]{Marois06}, which exploits
the rotation of the field of view on an altitude/azimuth telescope
with time to distinguish (and subtract) PSF features from astrophysical sources.
Marois' team observed the nearby (40 pc) A star HR 8799, which was
known to have an IR excess and to be relatively young (20--160 Myr), making
any orbiting planets likely to be bright in the near infrared.
\citet{Marois08}  reported the discovery of three faint companions to
HR 8799, with proper motions consistent with HR 8799 and detectable
orbital motion.   Comparison with Pleiades brown dwarf brightnesses demonstrated
that these objects had likely masses below 11 \mjup.
Further observations at Keck Observatory allowed \citet{Marois10} to
discover a fourth, $e$ component to the system.  The projected orbital
separations of the four planets range from 14--70 AU.  

This family of objects, the first imaged multi-planet system, poses
special challenges for planet formation theory.  \citet{Marois10}
argue that their masses and orbital radii (and the relative scarcity
of other systems such as this) are inconsistent with both
in situ gravitational-instability disk fragmentation and
core-accretion scenarios.  Like the close-in ``hot Jupiters'', these
distant planets would seem to implicate migration in a disk as a
primary architectural factor in planetary system formation.

\subsubsection{SPHERE, GPI, and Project 1640}

Several ``extreme'' adaptive optics systems with coronagraphs are in development or
operation, and will be capable of detecting young ($< 1$ Gyr) giant
planets at a few diffraction widths from the position of a bright
star.  Such systems employ deformable mirrors with several hundreds to
thousands of actuators, and typically observe bright stars in the near
infrared.  They use integral field spectrographs produce data cubes (i.e.\ a low resolution spectrum at
each angular position) and can exploit field rotation to employ a
variety of PSF subtraction and speckle suppression techniques.   They thus
produce large data volumes, including spectra of their imaged planets,
and will typically have narrow fully-corrected fields of view ($< 1^{\prime\prime}$).  

Project 1640 \citep{P1640}, already in operation, is a collaboration between the
American Museum of Natural History and the University of Cambridge.
It employs the PALM 3000 adaptive optics system on the Hale 
200-in telescope at Palomar Observatory.  SPHERE \citep{SPHERE} and
GPI \citep{GPI} are next-generation coronagraphic imagers on the VLT and Gemini South
telescopes, respectively.  Both will employ thousands of actuators and
execute campaigns to discover young Jupiter-mass planets orbiting at
several AU from the nearest Sun-like stars.  Depending on the adopted
planet formation and luminosity evolution model, the detection rate predicted for GPI
ranges between   10\% and 25\%, given a target sample with age $<$100 Myr within 75 pc
\citep{mcbride11a}.  Therefore, if these instruments observe $\sim$500 stars from this sample,
then there will be at least 50 new exoplanets discovered and characterized via direct imaging.

\subsection{Rocky and Habitable Worlds}
\subsubsection{HARPS, Keck/HIRES, and the Planet Finding Spectrograph}
The first RV-discovered planets had typical Doppler amplitudes of $\sim$
50--500 m/s; the 10 m/s barrier was breached several times
between 2000-2005, and detections between 2--5 m/s were common between
2005-2010.  The primary instruments making these detections were the
HIRES instrument on the Keck {\sc i} telescope (operated by various teams
of Marcy \& Butler) with precision as low as 1--3 m/s, and the HARPS
spectrograph (with its heritage from the Geneva team) which
regularly achieves precision below 1 m/s on bright stars.   Several
next generation planet finding spectrographs are being built or commissioned at
this writing, including the Planet Finding Spectrograph at Magellan,
HARPS-North, and ESPRESSO.  

There are two primary obstacles to further precision in radial
velocity surveys towards the 10 cm/s necessary for the detection of
true Earth analogs.  The first is an instrumental stability issue:
calibrating the wavelength solution of spectrographs to an order of
magnitude better precision than previously possible.  For
emission-lamp calibration, a fundamental limit is the lifetime and
stability of the thorium-argon lamps used as wavelength fiducial.  A
promising solution is the use of laser frequency combs, which provide
essentially arbitrary  levels of wavelength precision.  Such devices
will be used in HARPS-North and ESPRESSO.  Also of importance is
understanding and maximizing the consistency of the illumination of the output fibers
used to guide light to the spectrograph.  

For absorption-cell instruments, the practical limit is one's ability
to model the system, in particular the absorption cell, the slit
illumination function, the instrumental profile, and the star itself.
Progress here is primarily made through careful FTS scans of the
actual cell used at the telescope, improved modeling techniques that
account for scattered light, and better deconvolution techniques for
acquiring stellar templates.

The second obstacle is that of the fundamental stability of the stars
themselves.  Stars experience p-mode oscillations at the few m/s level
that must be either modeled or averaged over.  Stellar magnetic
surface activity also contributes to radial velocity signatures in
many ways, most importantly through rotationally modulated spots and
plage, and perhaps through long-term stellar cycles.  Modeling and
mitigating these effects, perhaps though careful use of
contemporaneous photometry, will be necessary to achieve the next
large improvement in RV precision for the next generation of planet
searches.

\subsubsection{Space-Based Transit Surveys}
\label{spacetransit}

Transit surveys for rocky planets around solar-type stars require
extremely precise relative photometry.  The fractional depth of the
transit of an Earth-sized planet passing in front of a solar radius
star is only $\delta \sim 8 \times 10^{-5}$. While relative photometry
at a few times this level has been achieved from the ground for
individual bright stars with specialized techniques (e.g.,
\citealt{johnson2009,southworth2009,colon2010}), obtaining $\la
10^{-4}$ relative photometry for the large ensembles of stars required
to detect numerous transiting systems is probably out of reach for
ground-based surveys, due to unavoidable systematics arising from
variations due to the Earth's atmosphere.  The stability afforded by
space-based surveys, on the other hand, enables relative photometry
for large numbers of stars that is limited primarily by photon and
astrophysical noise.  Furthermore, for space-based surveys it is
possible to obtain continuous photometry for very long periods of
time, without diurnal or weather interruptions.  This eliminates the
aliasing problems that are germane to single-site ground-based transit
surveys, and enables the detection of long period transiting systems,
which, due to the low duty cycles and long transit durations, are
extremely difficult to detect from the ground .
 
CoRoT is a CNES-led space mission with participation from ESA
and other international partners, with the primary goals of studying stars
via asteroseismology and detecting transiting planets \citep{baglin2003}.  The 27cm
telescope was launched in December 2006, and is located in a Low Earth Orbit.
CoRoT is equipped with a 3.05$^\circ$ by 2.7$^\circ$ camera that
primarily monitors fields in two different areas of the sky, located
toward Galactic longitudes of $\sim 40^\circ$ and $\sim 210^\circ$ \citep{auvergne2009}.
There are two dwell times for the fields; long fields are typically
observed for $\sim 150$ days, whereas short fields are observed for
$\sim 30$ days.  To date more than 130,000 stars in $\sim 20$ fields
have been monitored with a cadence of $8.5$ minutes \citep{michel2012}.  The stellar populations of these fields
vary dramatically, but anywhere from $40-60\%$ of the targets are
expected to be dwarf stars suitable for transit surveys \citep{cabrera2009,erikson2012}.  Over the
typical $R\sim 12-16$ magnitude range of the targets, CoRoT achieves
relative photometry on time scales of $\sim 2$ hours at the level of
$\sim 10^{-4}$ at the bright end, degrading to $\sim 10^{-3}$ at
$R\sim 16$ \citep{aigrain2009}.  The precision and cadence is sufficient to detect
Jupiter-sized companions over the entire magnitude range, whereas
Neptunes and Super-Earths can be detected around the brighter stars 
(e.g., \citealt{cabrera2009}).

To date, CoRoT has announced over 20 detections of transiting planets
and brown dwarfs.  Notable among these discoveries are
the detection of a transiting brown dwarf with a mass of $\sim 60~\mjup$ \citep{bouchy2011}, the detection
of a Jupiter-sized planet with a relatively long period of $\sim 95$ days \citep{deeg2010}, and
the first detection of a transiting Super-Earth, with a radius of $\sim 1.7R_\oplus$
and a mass of $3-10~M_\oplus$ \citep{leger2009}.

{\it Kepler} is a NASA mission launched in March of 2009, with the primary goal of measuring the
frequency of rocky planets in the Habitable Zones of sunlike stars
\citep{borucki2010}.  To accomplish this, the 0.95m {\it Kepler}
telescope situated in an Earth-trailing orbit is monitoring a 105
square degree field-of-view near the constellation Cygnus continuously
over the 7+ year lifetime of the mission.  Light curves
of $\sim 200,000$ stars have been obtained over the course of the
mission, with typical sampling cadences of $\sim 30$ minutes,
amounting to $\sim 24,000$ observations over the first $\sim 500$ days
of the mission for the subset of stars monitored continuously during
this time \citep{batalha2012}.  A subset of stars have higher cadences
of $\sim 1$ minute.  The majority of the target stars are solar-type
dwarfs with $T_{\rm eff} \sim 5000-6500~{\rm K}$ and $\log{g} \sim 4.5$,
but the full range of target properties span $T_{\rm eff}\simeq
3500-10,000~{\rm K}$ and $\log g \simeq 3-5$ \citep{batalha2012}.  The
photometric precision and intrinsic stellar variability of this
sample is discussed in \citet{gilliland2011}. Relevant to the primary
mission goal, the photometric variability for $V\la 12$ stars on
transit time scales is $\sim 3\times 10^{-5}$, quite an impressive
figure, but $\sim 50\%$ larger than originally anticipated,
primarily due to the fact that typical target stars turn out to be
a factor of $\sim 2$ times more variable than the Sun, probably due to
their relative youth.  Although this increased
noise reduces the expected sensitivity and so yield of the mission for
habitable Earthlike planets, this reduced sensitivity was offset 
with a mission extension \citep{gilliland2011}.

Although the data set is not yet sufficient to reliably detect true
Earth analogs, the exquisite photometric precision and large sample
size has already enabled an fantastic array of science. 
In particular, based on the first $\sim 500$ days of
data, a total of $\sim 2,300$ transiting planets candidates have been
identified \citep{batalha2012}, including $\sim 360$ multi-planet
systems \citep{fabrycky2012}.  The majority of these candidates
are smaller than Neptune.  Despite the fact that a small fraction of these
signals have been confirmed by either radial velocity or transiting timing
methods, the overwhelming majority of these signals are expected to be due to real
planets, from a number of lines of evidence (e.g.,
\citealt{morton2011,lissauer2012}).  Over 30 systems have been confirmed
from various methods, including a system with six transiting planets \citep{Kepler-11},
the first discovery of a circumbinary planet \citep{doyle2011},
and several planets with radius of $\la R_\oplus$ \citep{muirhead2012, fressin2012}.

The masses and semimajor axes of the confirmed {\it CoRoT} and {\it
Kepler} systems with mass measurements as of Dec.\ 2011 are shown in
Figure \ref{fig:exoplanets}.  Comparing these to the planets
discovered by ground-based transit surveys highlights the large
expansion of discovery space that is enabled by going to space.

\subsubsection{Second Generation Microlensing Surveys\label{sec:futuremicro}}

Microlensing exoplanet searches are currently in the midst of a
transition to the second generation of surveys that will enable the routine
detection of rocky planets.  Although there have
been substantial modifications and upgrades to the details of the
``two-tier'' survey strategy initially suggested by \citet{gl1992}, up
until very recently this basic approach has been in use.  Second
generation surveys will operate in a very different manner.  With the
development of very large format CCD cameras with fields of view of a
square degree or greater, it becomes possible to monitor tens of
square degrees of the Galactic bulge containing roughly 100 million
stars with cadences of tens of minutes.  These cadences are sufficient
to detect both the primary event {\it and} detect the perturbations
from low-mass ($\sim M_\oplus$) planets, therefore obviating the need
for a follow-up observations and allowing for more uniform data and a
more objective detection criteria.  In order to detect all the
planetary perturbations, including those that last less than a day, a
longitudinally-distributed network of $1-2$m class telescopes equipped
with such wide FOV cameras is required.

The transition to the next generation survey model began in 2006 when
MOA upgraded to the dedicated MOA-II telescope in New Zealand, which
has a diameter of 1.8m and 2.2 deg$^2$ FOV
\citep{sako2008,hearnshaw2006}.  In 2010, OGLE upgraded to the 1.4
deg$^2$ OGLE-IV camera on their dedicated 1.3m telescope in Chile
\citep{udalski2009}.  The Wise Observatory 1.0m telescope in Israel
recently been equipped with a 1 deg$^2$ camera \citep{gorbikov2010}.  These
three groups are collaborating to continuously monitor an $\sim 8$
deg$^2$ region of the bulge with cadences of 15-30 minutes
\citep{shvartzvald2012}, and the first planet detection with this strategy was
recently announced \citep{yee2012}.

The next milestone in the development of second generation
microlensing surveys will be the completion of the Korean Microlensing
Telescope Network (KMTNet).  KMTNet is an ambitious, fully funded plan
by the Korean government to build three identical 1.6m telescopes with
4 deg$^2$ FOV cameras.  These will be located in South America, South
Africa, and Australia.  First light for the final telescope is
scheduled for late 2014.  

These second generation surveys are expected in increase the planet yields
by roughly an order of magnitude over current surveys 
\citep{bennett2004,shvartzvald2012}, and enable the detection
of Earth-mass planets, as well as free-floating planets with masses
greater than $\sim 10~M_\oplus$.  

\section{Conclusions}

In the roughly two decades since the first detections of planets
outside the solar system, the field of exoplanets has grown
enormously, developing into one of the forefront research areas in
astronomy.  The count of confirmed planets is now over 700, with the
sample doubling in size every few years at the current rate.  New
techniques, methods, experiments, instruments, telescopes and
satellites to detect exoplanets are constantly being developed,
and are enabling the detection and characterization of an increasingly
broad diversity of planets orbiting a wider and wider range of hosts.
These efforts are not only continually uncovering new and unexpected
types of planetary systems, but are beginning to allow for the robust
statistical characterization of the demographics of large samples of
exoplanets spanning a wide range of parameter space.  These efforts
ultimately serve to allow us achieve the more general goals of placing
our solar system in the context of planetary systems through the
Galaxy, understanding the physics of planetary formation and
evolution, and determining the frequency of habitable and inhabited
worlds.

\section*{ACKNOWLEDGMENTS}

B.S.G. would like to thank Thomas Beatty and Karen Mogren for permission to discuss
results in advance of submission, and acknowledges support from NSF CAREER Grant AST-1056524. 

J.T.W.\ acknowledges support by funding from the Center for Exoplanets
and Habitable Worlds.  The Center for Exoplanets and Habitable Worlds
is supported by the Pennsylvania State University, the Eberly College
of Science, and the Pennsylvania Space Grant Consortium.

This work makes extensive use of NASA's Astrophysics Data System, the
Exoplanet Orbit Database at exoplanets.org, and the Extrasolar Planets Encyclopedia
at exoplanet.eu.

\end{document}